\documentclass[aps,prd,preprintnumbers,nofootinbib,superscriptaddress,
fleqn,floatfix,tightenlines,twocolumn,compress,10pt]{revtex4} 
\usepackage{amsmath,amsfonts,amssymb,amscd,amsxtra,amsthm}
\usepackage{graphicx}  
\usepackage{epstopdf}
\usepackage{dcolumn}  
\usepackage{bm}          
\usepackage{slashed}
\usepackage{cancel}
\usepackage{float}
\usepackage{mathtools}
\usepackage{amsbsy}
\usepackage{amstext}
\usepackage{tabularx}
\usepackage{enumitem}  
\usepackage{array} 
\usepackage{tikz}
\usepackage{multirow}
\usepackage{physics}
\usepackage{makecell}
\usepackage{subcaption}
\usepackage{dsfont}
\usepackage{hyperref}
\usepackage{mathrsfs} 
\usepackage{soul}

\usepackage[utf8]{inputenc} 
\usepackage{booktabs} 
\usepackage[normalem]{ulem} 

\soulregister{\cite}{7}
\soulregister{\ref}{7}
\setstcolor{red}

\makeatletter

\begin{document} 
\title{Relativistic spatial distributions of transverse angular momentum}

\author{C\'edric Lorc\'e}
\email[E-mail: ]{cedric.lorce@polytechnique.edu}
\affiliation{CPHT, CNRS, \'Ecole polytechnique, Institut Polytechnique de Paris, Palaiseau, France}

\author{Asmita Mukherjee}
\email[E-mail: ]{asmita@phy.iitb.ac.in}
\affiliation{Department of Physics,
Indian Institute of Technology Bombay, Powai, Mumbai, India}

\author{Ravi Singh}
\email[E-mail: ]{ravi.singh298@iitb.ac.in}
\affiliation{Department of Physics,
Indian Institute of Technology Bombay, Powai, Mumbai, India}

\author{Ho-Yeon Won}
\email[E-mail: ]{hoyeon.won@polytechnique.edu}
\affiliation{CPHT, CNRS, \'Ecole polytechnique, Institut Polytechnique de Paris, Palaiseau, France}

\date {\today}
\begin{abstract}

In our previous work~\cite{Lorce:2025pxt}, we investigated the 2D spatial distributions of transverse total angular momentum, including orbital angular momentum and intrinsic spin, relative to the canonical center (or center of spin). 
In the present work, we extend this analysis in two directions. 
First, we study the corresponding transverse boost distributions relative to the canonical center. 
Second, since the definition of generalized angular momentum density depends crucially on the choice of pivot, we analyze how the spatial distributions of transverse total angular momentum and boost are modified when they are defined relative to different relativistic centers, viz.~the relativistic centers of mass, energy, and spin.
Considering spin-$1/2$ targets, we derive the corresponding 2D spatial distributions in the transverse plane, and further investigate how the spatial patterns evolve under longitudinal Lorentz boosts.
Additionally, we provide the corresponding light-front distributions in the transverse plane, establishing a clear connection between the instant-form and light-front descriptions of transverse angular momentum and boost.
\end{abstract}
\maketitle
\section{Introduction}
Nucleons are one of the most basic composite particles made of quarks and gluons, yet they contain a plethora of interesting relativistic dynamics that underlie their seemingly simple property like spin~\cite{Leader:2013jra, Lorce:2018egm}. 
This rich internal structure, governed by principles of quantum chromodynamics (QCD), is encoded in the spatial distribution of orbital angular momentum (OAM), intrinsic spin, and energy dipole moment of quarks and gluons~\cite{Lorce:2017wkb,Mukherjee:2024xnu,Lorce:2025pxt}. 
While the longitudinal components of these quantities have been extensively studied, their transverse counterparts remain comparatively little explored due to their dependence on the hadron momentum and on the choice of pivot about which they are defined. A comprehensive relativistic analysis of transverse angular momentum (AM) and boost distributions, including their transformation properties and pivot dependence, is therefore essential for understanding how hadron spin emerges from quark and gluon degrees of freedom in position space. Such spatially resolved information is also of direct relevance to the physics program of the Electron–Ion Collider~\cite{Accardi:2012qut,Aschenauer:2017jsk,AbdulKhalek:2021gbh}, which aims to access multidimensional picture of hadron structure.

The definition of AM generally requires the choice of a pivot, i.e., the reference point about which the system rotates.
Conventionally, the origin of the coordinate system is chosen to coincide with the pivot, although this choice is arbitrary.
For a pointlike particle, the pivot is naturally identified with the particle position, and the spin is then defined as the total angular momentum (TAM) relative to this point.
For a composite system in classical mechanics, the center of mass provides a natural pivot.
In a relativistic quantum field-theoretic description, however, the notion of position is not unique, and the internal AM depends on the chosen pivot.
For the nucleon, three standard relativistic pivots are commonly considered: the relativistic centers of energy, mass, and spin~\cite{Bakker:2004ib,Burkardt:2002hr,Burkardt:2005hp,Lorce:2018zpf}.
This pivot dependence is particularly important for transverse TAM, since the transverse TAM operators do not commute with the generator of longitudinal Lorentz boosts.
This non-commutativity is the origin of longstanding debates on transverse spin sum rules~\cite{Bakker:2004ib,Leader:2011cr,Ji:2013tva,Harindranath:2013goa,Harindranath:2001rc,Ji:2020hii,Lorce:2021gxs}, which has recently been resolved by properly taking into account the role of the pivot~\cite{Lorce:2018zpf,Lorce:2021gxs}.

Spatial distributions of OAM and boost are obtained from Fourier transforms of matrix elements of the energy–momentum tensor (EMT) and are conventionally defined in two frames: the Breit frame (BF)~\cite{Polyakov:2002yz} in instant-form (IF) coordinates, and the Drell–Yan frame (DYF) in light-front (LF) coordinates~\cite{Lorce:2018egm,Freese:2021czn}. 
The BF allows one to define three-dimensional (3D) spatial distributions, but their interpretation as true densities is obscured by relativistic recoil effects~\cite{Kelly:2002if,Miller:2007uy,Jaffe:2020ebz}. 
In contrast, the DYF provides well-defined two-dimensional (2D) transverse densities due to the Galilean symmetry of the LF formalism, a feature that can also be realized in the infinite-momentum frame (IMF) within IF dynamics~\cite{Soper:1976jc,Burkardt:2000za, Miller:2010nz}. However, the rest-frame picture of the hadron in DYF is distorted due to switching to LF coordinates~\cite{Miller:2007uy,Carlson:2007xd}.
To interpolate between the BF and the IMF, the elastic frame (EF) was introduced within the quantum phase-space formalism, where relativistic spatial distributions are interpreted as quasi-densities~\cite{Lorce:2017wkb,Lorce:2018egm} in the Wigner sense. 
It also allows one to understand the LF distortions in terms of relativistic effects~\cite{Lorce:2020onh,Lorce:2022jyi,Chen:2022smg,Chen:2023dxp,Won:2025dgc,Won:2026ljg}. Defining transverse OAM distributions in EF is problematic, because transverse OAM involves derivatives with respect to the longitudinal momentum transfer $\Delta^z$, whereas the EF enforces $\Delta^z=0$. 
To overcome these limitations, we introduced in our earlier work, a generic frame (GF) in 3D space. Although the GF involves non-zero energy transfer and thus violates the elastic condition, integrating these 3D distributions over the longitudinal coordinate yields distributions in the plane transverse to the motion of the target, effectively projecting the GF onto the 2D EF and restoring elasticity $\Delta^0=0$.

In our previous work~\cite{Lorce:2025pxt}, we investigated the spatial distributions of transverse OAM, intrinsic spin, and TAM relative to the canonical center in the transverse plane, verifying the transverse spin sum rule relative to the canonical center.
In the present study, we extend this analysis in two directions: we include the transverse boost (or more precisely energy dipole moment) distribution relative to the canonical center in the transverse plane and investigate the 2D spatial distributions of transverse OAM, intrinsic spin, TAM, and boost for different choices of pivot. Using the spatial distributions of momentum and energy studied in~\cite{Won:2025dgc}, we explore the pivot dependence of transverse OAM and boost at the level of distributions, thus providing a comprehensive understanding of their spatial structure inside hadrons.
The current formalism also makes it possible to show how the spatial distributions of boost and OAM in the IF formalism give rise to the corresponding LF distributions, thus explaining the distortions displayed by the latter.

The paper is organized as follows.
In Sec.~II, we review the generalized AM operator density, including the OAM, intrinsic spin, TAM, and boost.
The matrix elements of the EMT and generalized intrinsic spin for spin-$1/2$ targets are also discussed.
In Sec.~III, we construct the relativistic spatial distributions in the transverse plane using the quantum phase-space formalism by integrating the 3D spatial distributions obtained in the generic frame.
In this framework, the internal angular momentum associated with the relativistic centers is defined and mapped onto the corresponding 2D spatial distribution in the IF formalism. We present numerical results for the nucleon, including the effects of the target momentum, polarization, and pivot choice.
In Sec.~IV, we formulate the corresponding LF distributions, discuss their spatial structure and establish the connection with the IF distributions in the IMF limit.
Finally, Sec.~V summarizes our main findings. Technical details and complementary discussions are collected in the Appendices.
In Appendix~\ref{app:external_TAM}, we formulate the spatial distributions of the external AM in the quantum phase-space formalism. In Appendix~\ref{app:pion}, we apply the formalism of transverse generalized AM distributions to the simpler case of a spin-$0$ target (pion) and, in Appendix~\ref{app_LF_genuine_spin_operator}, we review the genuine LF spin operator and its 2D spatial structure.

Throughout this work, products of noncommuting Hermitian operators are understood in the symmetrized form, e.g., $AB:=(AB+BA)/2$, in order to avoid ordering ambiguities~\cite{Pryce:1935ibt,Born:1935ap}.
Early Latin indices $a,b,c,\cdots$ run over the transverse spatial directions $1,2$, whereas middle Latin indices $i,j,k,\cdots$ run over all three spatial directions $1,2,3$.

\section{Angular momentum operators and matrix elements}
\subsection{Poincaré generators and energy-momentum tensor operators}

Poincar\'e symmetry in quantum field  theory (QFT) gives rise, via Noether's theorem, to conserved currents, viz. the EMT $\hat{T}^{\alpha\beta}(x)$ for spacetime translations, and the generalized AM tensor $\hat{J}^{\mu\alpha\beta}(x)$ for Lorentz transformations.
The generalized AM and four-momentum operators are then derived from the spatial integral of the corresponding Noether currents as follows
\begin{align}
    \hspace{-0.5cm}
    \hat{J}^{\alpha\beta}
& = \int_{\Sigma} d\Sigma_{\mu}\,
    \hat{J}^{\mu\alpha\beta}   (x), 
    \hspace{0.5cm}
    \hat{P}^{\nu}
  =  \int_{\Sigma} d\Sigma_{\mu}\,
    \hat{T}^{\mu\nu}   (x),
\label{def_AM}
\end{align}
where the time-dependence on the left-hand sides drops out due to the conservation laws $\partial_{\mu}\hat{J}^{\mu\alpha\beta}(x)=0$ and $\partial_{\mu}\hat{T}^{\mu\nu}(x)=0$, and to the standard assumption that surface terms do not contribute.
$\Sigma$ denotes the quantization hypersurface and $d\Sigma_{\mu}$ is the corresponding directed infinitesimal hypersurface element~\cite{Ji:1995ft,Ji:2001xd,Ji:2012ux,Ji:2014vha,Li:2015hew,Ji:2018pic,Ma:2021yqx,Ji:2026sns}. 
It can be written as $d\Sigma_{\mu}=n_{\mu}\,d\Sigma$, where $n^{\mu}$ specifies the normal direction to $\Sigma$.

The generalized AM tensor operator density 
\begin{align}
    \hat{J}^{\mu\alpha\beta}   (x)
& = \hat{L}^{\mu\alpha\beta}   (x)
  + \hat{S}^{\mu\alpha\beta}   (x),
\end{align}
is further decomposed into generalized OAM tensor $\hat{L}^{\mu\alpha\beta}(x)$ and generalized intrinsic spin tensor (GIST) $\hat{S}^{\mu\alpha\beta}(x)$ contributions defined as\footnote{The quark and gluon fields $\psi(x), \bar\psi(x), A^\mu(x)$, and gluon field-strength tensor $F^{\mu\nu}(x)$ are understood to be operator-valued quantum fields, and hats on these elementary field operators are omitted for notational simplicity.}
\begin{align}
    \hat{L}^{\mu\alpha\beta}
    (x)
&:= x^{\alpha} 
    \hat{T}^{\mu\beta } (x) 
  - x^{\beta } 
    \hat{T}^{\mu\alpha} (x), 
\label{def_OAM_operator_densities}\\
    \hat{S}^{\mu\alpha\beta}
    (x)
&:= \frac{1}{2}
    \epsilon^{\mu\alpha\beta\lambda}
    \bar{\psi}  (x)
    \gamma_{\lambda}\gamma_{5}
    \psi  (x),
\label{def_intrinsic_spin_operator_densities}
\end{align}
with $\epsilon_{0123}=1$.
Note that the latter only includes the quark degrees of freedom in the context of the present work. 

On the equal-time hypersurface at $x^{0}=\mathrm{constant}$, the TAM operators and boost generators are defined via the six independent components of the generalized AM operator as
\begin{align}
    \hat{J}^{i}
&:= \frac{1}{2}
    \epsilon^{ijk} \hat{J}^{jk},  
    \hspace{1cm}
    \hat{K}^{i}
 := \hat{J}^{0i}, 
\label{def_AM_operators}
\end{align}
which generate spatial rotations about the origin and Lorentz boosts with respect to the origin, respectively.
The TAM operators are given by the sum of the OAM and intrinsic spin contributions, i.e., $\hat{J}^{i}=\hat{L}^{i}+\hat{S}^{i}$, with each term defined analogously.

In QCD, the local gauge-invariant and asymmetric EMT current 
\begin{align}
    \hat{T}^{\mu\nu}
    (x)
& = \hat{T}_{q}^{\mu\nu}(x)
  + \hat{T}_{G}^{\mu\nu}(x),
\end{align}
is given by the sum of gauge-invariant quark and gluon contributions 
\begin{align}
    \hat{T}_{q}^{\mu\nu}
& = \bar{\psi} 
    \gamma^{\mu}    
    \frac{i}{2}  \overleftrightarrow{D}^{\nu} 
    \psi, \\
    \hat{T}_{G}^{\mu\nu}
& = 
    F^{a,\mu\lambda}  F_{\phantom{a,}\lambda}^{a,\phantom{\lambda}\nu}
  - \frac{1}{4}
    g^{\mu\nu}\,
    F^{a,\lambda\rho}  F_{\phantom{a,}\rho\lambda}^{a}.
\label{EMTcurrent}
\end{align}
$\overleftrightarrow{D}^{\mu}=\overrightarrow{\partial}^{\mu}-\overleftarrow{\partial}^{\mu}-2igA^{\mu}$ and 
$F^{a,\mu\nu}=\partial^{\mu}A^{a,\nu}
-\partial^{\nu}A^{a,\mu}+gf^{abc}A^{b,\mu}A^{c,\nu}$ denote the covariant derivative and gluon field-strength tensor, respectively.
The antisymmetric part of the quark EMT is connected to the axial-vector current through the QCD equation of motion~\cite{Lorce:2017wkb,Leader:2013jra}:
\begin{align}
    \hat{T}_{q}^{[\alpha\beta]}
    (x)
& = - \partial_\mu 
    \hat{S}^{\mu \alpha \beta}(x),
\label{antiEMT}
\end{align}
with the notation $V^{[\mu}W^{\nu]}:=V^{\mu}W^{\nu}-V^{\nu}W^{\mu}$.
Since the gluon TAM operator cannot be decomposed in a local and gauge-invariant manner into OAM and intrinsic spin contributions, there is no gluon contribution to the GIST operator density~\cite{Leader:2013jra}.

\subsection{Matrix elements and parametrizations} 

The matrix elements of some operator $\hat{O}(0)$ for spin-1/2 targets with mass $M$ can be parameterized as follows
\begin{align}
    \mel{p^{\prime},s^{\prime}}{\hat{O}(0)}{p,s}
& = \bar{u}(p,s)
    \Gamma[\hat{O}]
    u(p,s).
\end{align}
The four-momentum states are normalized as $\braket{p^{\prime},s^{\prime}}{p,s}=2p^{0}
\left(2\pi\right)^{3}\delta^{(3)}\left(\bm{p}^{\prime}-\bm{p}\right)\delta_{s^{\prime}s}$ with the canonical spin polarizations $s^{\prime},s$.
The normalization of Dirac spinors is $\bar{u}(p,s)u(p,s)=2M$. 
With the average momentum $P=(p^{\prime}+p)/2$ and momentum transfer $\Delta=p^{\prime}-p$, the on-shell conditions $p^{2}=p^{\prime 2}=M^{2}$ imply $P\cdot\Delta=0$ and $P^{2}=M^{2}(1+\tau)$ with the dimensionless Lorentz invariant variable $\tau:=-\Delta^{2}/(4M^{2})$.

The quark EMT vertex is given by~\cite{Bakker:2004ib,Leader:2013jra,Lorce:2017wkb,Burkert:2023wzr}
\begin{align}
   \hspace{-.7cm} \Gamma[\hat{T}_{q}^{\mu\nu}]
& = \frac{P^{\mu}P^{\nu}}{M}  
    A^{q}
  + \frac{\Delta^{\mu}\Delta^{\nu}-g^{\mu\nu}\Delta^{2}}{4M}  
    D^{q}+ M g^{\mu\nu} 
    \bar{C}^{q} \notag\\
&
  + \frac{iP^{\{\mu}\sigma^{\nu\}\rho}\Delta_{\rho}}{2M}  
    J^{q}- \frac{iP^{[\mu}\sigma^{\nu]\rho}\Delta_{\rho}}{2M}  
    S^{q}, 
    \label{EMT}
\end{align}
with the notation $V^{\{\mu}W^{\nu\}}:=V^{\mu}W^{\nu}+V^{\nu}W^{\mu}$ and a similar expression for the gluon EMT. For the GIST operator, we have
\begin{equation} 
    \Gamma[\hat{S}^{\mu\alpha\beta}]
 = \frac{1}{2} \epsilon^{\mu\alpha\beta\lambda}
    \left(
    \gamma_{\lambda} \gamma_{5}
    G_{A}^{q}  
  + \frac{\Delta_{\lambda}\gamma_{5}}{2M}  
    G_{P}^{q}   
    \right),
\label{axial}
\end{equation}
For simplicity, we will often omit the parton-type label in the following and specify it only when necessary. 
In addition, the coefficients, called form factors (FFs), are real-valued functions of the squared four-momentum transfer, $t=\Delta^{2}$. 

Due to Poincaré symmetry, the total EMT FFs must satisfy the constraints~\cite{Ji:1996ek,Leader:2013jra,Teryaev:1999su,Lowdon:2017idv,Cotogno:2019xcl}
\begin{alignat}{3}
    A(0)
& = A^{q}(0)
& + A^{G}(0)
& = 1, \notag\\
    J(0)
& = J^{q}(0)
& + J^{G}(0)
& = \frac{1}{2}, \notag\\
    \bar{C}(t)
& = \bar{C}^{q}(t)
& + \bar{C}^{G}(t)
& = 0.
    \label{GFFconstraints}
\end{alignat}
Note that the individual quark and gluon contributions to the FFs do not in general satisfy these constraints independently.
The value of $D(0)$ is not constrained by Poincar\'e symmetry, but, based on stability arguments, it has been conjectured to be negative in QCD~\cite{Burkert:2023wzr,Perevalova:2016dln,Polyakov:2018zvc,Lorce:2025oot}.

Since the GIST operator in Eq.~\eqref{def_intrinsic_spin_operator_densities} is constructed solely from quark fields, the QCD equation of motion in Eq.~\eqref{antiEMT} relates the quark intrinsic spin FF to axial-vector FF, while the corresponding gluon FF vanishes~\cite{Lorce:2017wkb}
\begin{align}
    S^{q}(t)
  = \frac{1}{2}
    G_{A}^{q}(t) ,
    \qquad
    S^{G}(t)
  = 0.
\label{spinrelation}
\end{align}

For the numerical illustrations presented in Secs.~\ref{numerical_discussion_IF} and \ref{numerical_discussion_LF},  we adopt a simple multipole parametrization of the relevant FFs. 
In particular, for the nucleon (see Appendix~\ref{app:pion} for the pion) we use the tripole ansatz for the EMT FFs based on Ref.~\cite{Hackett:2023rif}.\footnote{For the nucleon, the tripole mass $\Lambda_{\mathcal{F}_a}$ is obtained by multiplying the corresponding dipole mass by $\sqrt{3/2}$, such that $\left.d\mathcal{F}_a(t)/dt\right|{t=0}$ remains unchanged~\cite{Lorce:2018egm}.} 
This parametrization yields spatial distributions consistent with the momentum and energy sum rules. Unless otherwise stated, the transverse AM distributions are displayed only up to $P^{z}=10~\mathrm{GeV}$, since they diverge in the limit $P^{z}\to\infty$.

\section{Relativistic spatial distributions}

We first introduce a generic frame to define the 3D relativistic spatial distribution of transverse OAM~\cite{Lorce:2025ll}.
We then obtain the corresponding 2D distribution in the transverse plane by integrating the 3D distribution over the longitudinal coordinate.

\subsection{Generic frame}
The 3D GF is a class of Lorentz frames with nonzero target momentum, which provides a continuous interpolation between the BF and the IMF.
In this frame, the average momentum $P$ and momentum transfer $\Delta$ are given by 
\begin{align}
    P^{\mu}
& = \left(P^{0},\bm{0}_{\perp},P^{z}\right), 
    \hspace{0.5cm}
    \Delta^{\mu}
  = \left(\Delta^{0},\bm{\Delta}_{\perp},\Delta^{z}\right),
\end{align}
where we set $\bm{P}=P^{z}\bm{e}_{z}$ for simplicity.
From the on-shell condition
\begin{align}
    P\cdot\Delta=0
    \quad
    \Rightarrow
    \quad
    \Delta^{0}
  = \frac{\bm{P}\cdot\bm{\Delta}}{P^{0}},
\end{align}
we see that the energy transfer is in general nonzero in the GF. 
This implies that the initial and final states are associated with different Lorentz boost factors.

For a target with nonzero spin $j$, the Lorentz transformation of the matrix element of a rank-$n$ operator $\hat{O}^{\mu_{1}\mu_{2}\cdots\mu_{n}}$ involves both the usual tensor transformation and the Wigner spin rotation induced by the non-commutativity of boosts, $[\hat{K}^{i},\hat{K}^{j}]=-i\epsilon^{ijk}\hat{J}^{k}$, as shown below~\cite{Jacob:1959at,Durand:1962zza}
\begin{align}
&   \hspace{-0.7cm}
    \mel{p^{\prime},s^{\prime}}{\hat{O}^{\mu_{1}\cdots\mu_{n}}(0)}{p,s}_{\mathrm{GF}}\notag\\
&   \hspace{-0.7cm}
  = \sum_{s_{\mathrm{BF}}^{\prime},s_{\mathrm{BF}}}
    D_{s_{\mathrm{BF}}s}^{(j)}
    (p_{\mathrm{BF}},\Lambda)\,
    D_{s_{\mathrm{BF}}^{\prime}s^{\prime}}^{*(j)}    
    (p_{\mathrm{BF}}^{\prime},\Lambda) \notag\\
&   \hspace{-0.7cm}
    \times    
    \Lambda_{\phantom{\mu_{1}}\nu_{1}}^{\mu_{1}}
    \cdots
    \Lambda_{\phantom{\mu_{n}}\nu_{n}}^{\mu_{n}}\!
    \mel{p_{\mathrm{BF}}^{\prime},s_{\mathrm{BF}}^{\prime}}
    {\hat{O}^{\nu_{1}\cdots\nu_{n}}\left(0\right)}
    {p_{\mathrm{BF}},s_{\mathrm{BF}}},
\label{LT}
\end{align}
where $D^{(j)}$ denotes the Wigner rotation matrix for spin-$j$ targets.
The Wigner rotation matrix for spin-1/2 targets is given by
\begin{align}
    D_{s^{\prime}s}^{(1/2)}
    (p,\Lambda)
  = \cos{\frac{\theta}{2}}\, \delta_{s^{\prime}s}
  + i \sin{\frac{\theta}{2}}\,
    \frac{\left(\bm{p}\times\bm{\sigma}_{s^{\prime}s}\right)^{z}}{\left|\bm{p}_{\perp}\right|},
\end{align}
with the Pauli spin matrix elements $\bm{\sigma}_{s^{\prime}s}$.
In general, there is no reason for the Wigner rotation angles associated with the initial and final states, denoted by $\theta$ and $\theta^{\prime}$, to coincide.

The projection $\Delta^{z}=0$ reduces the 3D GF to the 2D EF.
Through the on-shell condition $P\cdot\Delta=0$, it also enforces $\Delta^{0}=0$, so that the Wigner rotation angles associated with the initial and final states coincide, i.e., $\theta=\theta^{\prime}$.
Using the Dirac bilinear relations~\cite{Lorce:2017isp}, we calculate both the EF and BF matrix elements in Eq.~\eqref{LT} in the transverse plane and obtain the corresponding Wigner rotation angle and Lorentz boost factors~\cite{Lorce:2022jyi,Chen:2022smg,Chen:2024oxx,Won:2025dgc,Won:2026ljg}
\begin{align}
    \cos{\theta}
& = \frac{P^{0}+M\left(1+\tau\right)}{\left(P^{0}+M\right)\sqrt{1+\tau}},  \notag\\
    \sin{\theta}
& = - \frac{\sqrt{\tau}P^{z}}{\left(P^{0}+M\right)\sqrt{1+\tau}},
\label{WR}
\end{align}
and 
\begin{align}
    \gamma 
  = \frac{P^{0}}{\sqrt{P^{2}}}
  = \frac{P^{0}}{P_{\mathrm{BF}}^{0}}
  = \frac{P^{0}}{M\sqrt{1+\tau}},
    \qquad
    \beta
  = \frac{P^{z}}{P^{0}}.
\end{align}
As expected, these expressions are independent of the spin polarization and the form of the operator in the matrix element.

\subsection{Quantum phase-space formalism }
\label{QPSF} 

When calculating matrix elements of the form  $\mel{p^{\prime},s^{\prime}}{\int d^{3}x\, x^{i} \hat{O}(x)}{p,s}$ in defining relativistic spatial distributions of the AM, one encounters a divergent term, known to be associated with the contribution from the center of the wave-packet~\cite{Jaffe:1989jz,Bakker:2004ib,Lorce:2021gxs}. 
To discard this contribution and define the internal distribution in a systematic manner, we adopt the quantum phase‑space formalism, used to study the relativistic spatial distributions of the transverse TAM~\cite{Lorce:2025pxt}, the EMT~\cite{Lorce:2018egm,Won:2025dgc,Won:2026ljg,Lee:2026gxb}, the electromagnetic properties~\cite{Lorce:2020onh,Kim:2021kum,Lorce:2022jyi,Chen:2022smg,Kim:2022wkc,Chen:2023dxp,Hong:2023tkv}, and the axial-vector current~\cite{Chen:2024oxx,Chen:2024ksq} in the transverse plane for a moving hadron.

In the quantum phase-space approach, the hadronic wave-packet in hadronic matrix elements is factored out into a Wigner distribution (see Ref.~\cite{Chen:2023dxp} for details). 
The internal phase-space amplitude associated with some operator $\hat{O}$ is then given by the following Fourier transform
\begin{align}
    \langle \hat{O}\left(\bm{x}\right)\rangle_{\bm{\mathcal{R}},\bm{P}}^{s^{\prime}s}
& = \int \frac{d^{3}\Delta}{\left(2\pi\right)^{3}}\;
    e^{-i\bm{\Delta}\cdot(\bm{x}-\bm{\mathcal{R}})}
    \langle\langle \hat{O}\rangle\rangle  
\label{def_3D_internal_distribution}
\end{align}
with the shorthand
\begin{align}
    \langle\langle \hat{O}\rangle\rangle
 := \frac{\mel{p^{\prime},s^{\prime}}
    {\hat{O}(0)}
    {p,s}}{2\sqrt{p^{\prime0}p^{0}}}.
\end{align}
The quantity $\langle \hat{O}\left(\bm{x}\right)\rangle_{\bm{\mathcal{R}},\bm{P}}^{s^{\prime}s}$ can be interpreted as the internal distribution for a system localized around the (average) position $\bm{\mathcal{R}}$ and the (average) momentum $\bm{P}$ in the Wigner sense.
Using translation symmetry, one further finds that
\begin{align}
    \langle \hat{O}\left(\bm{x}\right) \rangle_{\bm{\mathcal{R}},\bm{P}}^{s^{\prime}s} 
& = \langle \hat{O}\left(\bm{x}-\bm{\mathcal{R}}\right) \rangle_{\bm{0},\bm{P}}^{s^{\prime}s},
\end{align}
so that the same quantity is equivalently viewed as the internal distribution at the relative position $\bm{r}=\bm{x}-\bm{\mathcal{R}}$ for a system localized around the origin of the coordinates.

\subsection{Internal angular momentum}

The definition of the AM operator generally depends on the center of rotation, i.e., pivot. 
The spin of a composite system in QFT corresponds to its internal AM defined relative to a specific relativistic pivot~\cite{Lorce:2021gxs}.

In QFT, the relativistic pivot is generally defined through three distinct position operators~\cite{PhysRev.137.B188,Pryce:1935ibt,Pryce:1948pf,Mller1949-MLLOTD,Choi:2014nea,Lorce:2018zpf,Lorce:2021gxs}, namely, the center of energy $\hat{R}_{E}^{\mu}$, the center of mass $\hat{R}_{M}^{\mu}$, and the canonical center $\hat{R}_{c}^{\mu}$
\begin{align}
    \hat{R}_{E}^{\mu}
&:= x^{0}
    \frac{\hat{P}^{\mu}}{\hat{P}^{0}}
  - \frac{\hat{J}^{0\mu}}{\hat{P}^{0}},\\
    \hat{R}_{M}^{\mu}
&:= \Lambda_{\phantom{\mu}\nu}^{\mu}
    \hat{R}_{E}^{\nu}|_{\mathrm{rest}},\\
    \hat{R}_{c}^{\mu}
&:= \frac{\hat{P}^{0}\hat{R}_{E}^{\mu}+\hat{M} \hat{R}_{M}^{\mu}}{\hat{P}^{0}+\hat{M}},
\end{align}
where $\hat{M}=M \hat{I}$ with $\hat I$ the identity operator. It has been shown that only the canonical position operator $\bm{\hat{R}}_{c}$ has mutually commuting components, and hence admits localized position eigenstates in the Newton-Wigner sense~\cite{Pryce:1948pf,Newton:1949cq,Foldy:1949wa,Lorce:2021gxs}:
\begin{align}
    \bm{\hat{R}}_{c}\ket{\bm{x}}
  = \bm{x}\ket{\bm{x}}.
\end{align}

Using these relativistic centers as pivots, one defines the corresponding internal AM and boost operators as~\cite{Lorce:2021gxs}
\begin{align}
    \bm{\hat{J}}_{X}
& = \bm{\hat{J}}
  - \bm{\hat{R}}_{X} \times \bm{\hat{P}}, 
    \\
    \bm{\hat{K}}_{X}
& = \bm{\hat{K}}
  - \hat{R}_{X}^{0}
        \bm{\hat{P}}
  + \bm{\hat{R}}_{X} \hat{P}^{0}.
\end{align}
In the rest frame, i.e. $\bm{P}=\bm{0}$, the three relativistic centers coincide, and so do the corresponding internal AM operators too.
They therefore all satisfy the $su(2)$ algebra.
For a moving system, $\bm{P}\neq\bm{0}$, however, the three centers are no longer identical.
In this case, only the internal AM operator defined relative to the canonical center $\hat{J}_{c}$ satisfies the $su(2)$ algebra and can therefore be interpreted as a genuine spin operator.
In this context, the canonical position operator has alternatively been referred to as the relativistic center of spin operator~\cite{PhysRev.137.B188,Lorce:2018zpf,Schwartz:2020lys}.
Although the internal AM operators defined relative to the centers of energy and mass do not satisfy the $su(2)$ algebra for $\bm{P}\neq\bm{0}$, we will still refer to them as spin operators in the following for convenience.

\subsection{Spatial distributions of internal angular momentum}

Within the quantum phase-space formalism, the relativistic centers are evaluated via $\bm{R}_{X}(s^{\prime},s):=\langle \hat{\bm{R}}_{X} \rangle_{\bm{\mathcal{R}},\bm{P}}^{s^{\prime}s}$~\cite{Lorce:2021gxs},
\begin{align}
    \bm{R}_{E}(s^{\prime},s)
& = \delta_{s^{\prime}s}
    \bm{\mathcal{R}}
  + \frac{\bm{P}\times\bm{\sigma}_{s^{\prime}s}}
    {2E_{P}\left(E_{P}+M\right)} , 
    \label{RE}\\
    \bm{R}_{M}(s^{\prime},s)
& = \delta_{s^{\prime}s}
    \bm{\mathcal{R}}
  - \frac{\bm{P}\times\bm{\sigma}_{s^{\prime}s}}
    {2M\left(E_{P}+M\right)}, 
    \label{RM} \\
    \bm{R}_{c}(s^{\prime},s)
& = \delta_{s^{\prime}s}
    \bm{\mathcal{R}},
\label{Rc}
\end{align}
with $E_{P}=\sqrt{(P^z)^{2}+M^{2}}$.
For convenience, we choose the origin of coordinates to be $\bm{\mathcal{R}}=\bm{0}$ to eliminate the wave-packet contribution mentioned above.
Therefore, the relativistic spatial distributions defined relative to the origin in the Wigner sense naturally correspond to the internal distributions of the system with respect to the canonical center~\cite{Jaffe:1989jz,Shore:1999be,Bakker:2004ib,Leader:2013jra,Lorce:2021gxs}.

In Eqs.~\eqref{RE}-\eqref{Rc}, the longitudinal components all coincide with each other.
Thus, only the transverse components of AM depend on the choice of the pivot. Their spatial distributions, defined relative to a relativistic center $\bm{R}_{X}$ at $x^0=0$, are given in the transverse plane by
\begin{align}
    \hspace{-0.7cm} 
    J_{X}^{a}   (\bm{b}_{\perp},P^{z};s^{\prime},s)
& = J_{c}^{a}   (\bm{b}_{\perp},P^{z};s^{\prime},s)  \notag\\
&   \hspace{-1cm}
  - \frac{1}{2}
    \epsilon_{\perp}^{ab}
    \sum_{s^{\prime\prime}}
    R_{X}^{b}(s^{\prime},s^{\prime\prime})
    P^{3}(\bm{b}_{\perp},P^{z};s^{\prime\prime},s) \notag\\
&   \hspace{-1cm} 
  - \frac{1}{2}
    \epsilon_{\perp}^{ab}
    \sum_{s^{\prime\prime}}
    P^{3}(\bm{b}_{\perp},P^{z};s^{\prime},s^{\prime\prime})
    R_{X}^{b}(s^{\prime\prime},s),
    \label{def_internal_am_1}\\
    \hspace{-0.7cm} 
    K_{X}^{a}   (\bm{b}_{\perp},P^{z};s^{\prime},s)
& = K_{c}^{a}   (\bm{b}_{\perp},P^{z};s^{\prime},s)  \notag\\
&   \hspace{-1cm}
  + \frac{1}{2}
    \sum_{s''}
    R_{X}^{a} (s^{\prime},s^{\prime\prime})
    E   (\bm{b}_{\perp},P^{z};s^{\prime\prime},s) \notag\\
&   \hspace{-1cm}
  + \frac{1}{2}
    \sum_{s''}
    E (\bm{b}_{\perp},P^{z};s^{\prime},s^{\prime\prime})
    R_{X}^{a}   (s^{\prime\prime},s),
    \label{def_internal_am_2}
\end{align}
where the vector $\bm{b}_{\perp}$ is the transverse part of the relative position vector $\bm{r}=(\bm{b}_{\perp},r^{z})$.

The last two terms above represent the subtraction of the contributions from the external AM, composed of the corresponding position vector and the relativistic spatial distributions of energy $E(\bm{b}_{\perp})$ and longitudinal momentum $P^{z}(\bm{b}_{\perp})$ given in~\cite{Won:2025dgc}.
The sum over spin polarization $s^{\prime\prime}$ arises from the nucleon projector part of the completeness relation inserted between the two operators in the quantum phase-space formalism, see Appendix~\ref{app:external_TAM} for details.

Since we set $\bm{\mathcal{R}}=\bm{0}$, thereby identifying the canonical center with the origin of the coordinate system, the transverse TAM $J_c^a$ and boost $K_c^a$ distributions relative to the canonical center are given by
\begin{align}
    J_{c}^{a}
    (\bm{b_\perp},P^{z};s^{\prime},s)
& = L_{c}^{a}
    (\bm{b_\perp},P^{z};s^{\prime},s) \notag\\
& + S_{c}^{a}
(\bm{b_\perp},P^{z};s^{\prime},s), 
    \label{JcanEF}\\
    K_{c}^{a}
(\bm{b_\perp},P^{z};s^{\prime},s)
& = - \int dr^{z}\, r^{a}
    \langle \hat{T}^{00}(\bm{r}) \rangle_{\bm{0},\bm{P}}^{s^{\prime}s},
    \label{KcanEF}
\end{align}
where
\begin{align}
    \hspace{-0.5cm}
    L_{c}^{a}
(\bm{b_\perp},P^{z};s^{\prime},s)
& = \epsilon^{ajk} 
    \int dr^{z}\, r^{j}
    \langle \hat{T}^{0k}(\bm{r}) \rangle_{\bm{0},\bm{P}}^{s^{\prime}s},
    \label{LcanEF}\\
    \hspace{-0.5cm}
    S_{c}^{a}
(\bm{b_\perp},P^{z};s^{\prime},s)
& = \frac{1}{2}
    \epsilon^{ajk}
    \int dr^{z}\, 
    \langle \hat{S}^{0jk}(\bm{r}) \rangle_{\bm{0},\bm{P}}^{s^{\prime}s}.
    \label{ScanEF}
\end{align}
Referring to the previous work~\cite{Lorce:2025pxt}, one can obtain the transverse TAM distribution defined relative to the canonical center of the nucleon in the transverse plane as
\begin{align}
&   J_{c}^{a} 
   (\bm{b}_{\perp},P^{z};s^{\prime},s) 
  = \int \frac{d^{2}\Delta_{\perp}}{\left(2\pi\right)^{2}}
    e^{-i\bm{\Delta}_{\perp}\cdot\bm{b}_{\perp}}\notag\\
&   \times
    \Bigl[
    \delta_{s^{\prime}s} 
    i   \epsilon^{ab}_\perp   
    \sqrt{\tau}
    X_{1}^{b}  
    \tilde{J}_{1}+ \sigma_{s^{\prime}s}^{a}
    \tilde{J}_{0} 
  + \sigma_{s^{\prime}s}^{b}
    \tau 
    X_{2}^{ba} 
    \tilde{J}_{2}
    \Bigr],
\label{Jk_spin1/2}
\end{align}
where we introduced the irreducible multipole tensors in 2D momentum-transfer space, $X_{1}^{a}=\Delta_{\perp}^{a}/|\bm{\Delta}_{\perp}|$, and $X_{2}^{ab}=\Delta_{\perp}^{a}\Delta_{\perp}^{b}/|\bm{\Delta}_{\perp}|^{2}-\delta_{\perp}^{ab}/2$~\cite{Kim:2022bia,Kim:2022wkc,Hong:2023tkv}.
The corresponding multipole amplitudes, i.e., $\tilde{J}_{n}$, are given in terms of FFs and Wigner rotation angle as 
\begin{align}
    \hspace{-0.8cm}
    \tilde{J}_{0}
& = - 2MP^{z}
    \frac{d}{dt}    
    \biggl[
    \frac{P^{0}}{M}
    \frac{1}{\gamma}
    \frac{\sin{\theta}}{\sqrt{\tau}}
    \tau
    (
    A 
  - 2 J
    )
    \biggr]\notag\\
    \hspace{-0.8cm}
& 
  - \frac{\beta}{\gamma}
    \frac{\sin{\theta}}{\sqrt{\tau}}
    \frac{\tau}{4}
    D 
  + \frac{\cos{\theta}}{\gamma}
    \frac{1}{2}
    \biggl(
    J
  - S
  + \frac{1}{2}
    G_{A} 
    \biggr)\notag\\
    \hspace{-0.8cm}
& + \frac{M}
    {4P^{0}}
    (
    G_{A} 
  - \tau        G_{P}
    )\notag\\
    \hspace{-0.8cm}
& - 2M^{2}
    \frac{d}{dt}
    \biggl[
    2\tau
    J
  - \frac{\cos{\theta}}{\gamma}
    \tau
    (
    J
  + S 
    )
    \biggr],
    \label{JT0}\\
    \hspace{-0.8cm}
    \tilde{J}_{1}
& = 4MP^{z}
    \frac{d}{dt}    
    \biggl(
    \frac{P^{0}}{M}
    \frac{\cos{\theta}}{\gamma}A 
  + \frac{1}{\beta\gamma}
    \frac{\sin{\theta}}{\sqrt{\tau}}
    2 \tau
    J 
    \biggr)  \notag\\
    \hspace{-0.8cm}
& + \beta
    \frac{\cos{\theta}}{\gamma}
    \frac{1}{2}
    D 
  + \frac{1}{\gamma}
    \frac{\sin{\theta}}{\sqrt{\tau}}
    \frac{1}{2} 
    (
    J
  - S
  + G_{A}
    )\notag\\
    \hspace{-0.8cm}
& + 4M^{2}
    \frac{d}{dt}    
    \left[
    \frac{1}{\gamma}
    \frac{\sin{\theta}}{\sqrt{\tau}}
    \tau
    \left(
    J
  + S
    \right) 
    \right],
    \label{JU1}\\
    \hspace{-0.8cm}
    \tilde{J}_{2}
& = 4MP^{z}
    \frac{d}{dt}    
    \biggl[
    \frac{P^{0}}{M}
    \frac{1}{\gamma}
    \frac{\sin{\theta}}{\sqrt{\tau}}
    (
    A 
  - 2 
    J 
    )
    \biggr] \notag\\
    \hspace{-0.8cm}
& + \frac{\beta}{\gamma}
    \frac{\sin{\theta}}{\sqrt{\tau}}
    \frac{1}{2}
    D
  + \frac{M}{2P^{0}}
    \biggl(
    \frac{1}{\beta\gamma}
    \frac{\sin{\theta}}{\sqrt{\tau}}
    G_{A}
  - G_{P}
    \biggr)\notag\\
    \hspace{-0.8cm}
& + 4M^{2}
    \frac{d}{dt}  
    \biggl[
    2
    J
  - \frac{\cos{\theta}}{\gamma}
    (
    J
  + S
    )
    \biggr]
    \label{JT2}.
\end{align}
The decomposition of the transverse TAM into OAM and intrinsic spin distributions in the transverse plane is discussed in Ref.~\cite{Lorce:2025pxt}.

Analogously, the relativistic spatial distribution of transverse boost defined relative to the canonical center of the nucleon in the transverse plane is derived using Eq.~\eqref{KcanEF}
\begin{align}
&   \hspace{-0.6cm}K_{c}^{a} 
    (\bm{b}_{\perp},P^{z};s^{\prime},s) 
  = \int \frac{d^{2}\Delta_{\perp}}{\left(2\pi\right)^{2}}
    e^{-i\bm{\Delta}_{\perp}\cdot\bm{b}_{\perp}}\notag\\
&   \hspace{-0.6cm}
    \times
    \Bigl[ 
    \delta_{s^{\prime}s}
    i \sqrt{\tau}
    X_{1}^{a}   
    \tilde{K}_{1} 
  + \epsilon_{\perp}^{ab}
    \sigma_{s^{\prime}s}^{b}
    \tilde{K}_{0}
  + \epsilon_{\perp}^{bc}
    \sigma_{s^{\prime}s}^{b}
    \tau
    X_{2}^{ca} 
    \tilde{K}_{2}
    \Bigr],
\label{Kk_spin1/2}
\end{align}
whose multipole structure is similar to that of the TAM, albeit with an extra Levi-Civita accounting for the difference of behavior under parity.
The corresponding multipole amplitudes in the above are given as 
\begin{align}
    \tilde{K}_{0} 
& = - 
    2MP^{z}
    \frac{d}{dt}
    \Bigg[
    \frac{P^{0}}{M}
    \frac{1}{\beta\gamma}
    \frac{\sin{\theta}}{\sqrt{\tau}}
    \tau    
    A   \notag\\
& + \tau
    J   
  - \frac{M}{2P^{z}}
    \frac{1}{\gamma}
    \frac{\sin{\theta}}{\sqrt{\tau}}
    \tau
    F
    \Bigg],
    \label{KT_0} \\
    \tilde{K}_{1}
& = 4M^{2}
    \frac{d}{dt}
    \Bigg[
  - \frac{\left(P^{0}\right)^{2}}{M^{2}}
    \frac{\cos{\theta}}{\gamma}
    A     \notag\\
& + 2
    \tau
    J     
  + \frac{\cos{\theta}}{\gamma}
    F
    \Bigg]
    \label{KT_1},\\
    \tilde{K}_{2}
& = 4MP^{z}
    \frac{d}{dt}
    \Bigg[
    \frac{P^{0}}{M}
    \frac{1}{\beta\gamma}
    \frac{\sin{\theta}}{\sqrt{\tau}}
    A   \notag\\
& + 2
    J
  - \frac{M}{P^{z}}
    \frac{1}{\gamma}
    \frac{\sin{\theta}}{\sqrt{\tau}}
    F
    \Bigg],
    \label{KT_2}
\end{align}
with $F=-\bar{C}-\tau D$.
In the transverse plane, these spatial distributions contain no contribution proportional to the longitudinal spin polarization, $\sigma_{s^{\prime}s}^{z}$, since such terms arise from the time-dependent part and vanish at $r^{0}=0$.
Thus, the multipole structure further shows that the transverse AM distributions at $r^{0}=0$ in the transverse plane are identical for unpolarized and longitudinally polarized nucleons.

\subsection{Sum rules: transverse spin and boost}

Integrating the transverse TAM distribution in Eq.~\eqref{def_internal_am_1} with the relativistic position vectors in Eqs.~\eqref{RE}-\eqref{Rc} over the transverse plane
\begin{align}
    J_{X}^{a}
    (P^{z};s^{\prime},s)
& = \int d^{2}b_{\perp}\,
    J_{X}^{a}
    (\bm{b}_{\perp},P^{z};s^{\prime},s),
\end{align}
and using the FF relations in Eqs.~\eqref{GFFconstraints} and \eqref{spinrelation}, and the normalization of longitudinal momentum distribution~\cite{Won:2025dgc}, we obtain
\begin{align}
    J_{E}^{a}
    (P^{z};s',s)
& = \frac{M}{E_{P}} \frac{\sigma_{s^{\prime}s}^{a}}{2},
    \label{spin_sum_rule_energy}\\
    J_{M}^{a}
    (P^z;s',s)
& = \frac{E_{P}}{M} \frac{\sigma_{s^{\prime}s}^{a}}{2}.
    \label{spin_sum_rule_mass} \\
    J_{c}^{a}
    \left(P^{z};s^{\prime},s\right)
& = \frac{\sigma_{s^{\prime}s}^{a}}{2}.
    \label{spin_sum_rule_canonical} 
\end{align}
These results agree with the transverse spin sum rules defined relative to the relativistic centers of energy, mass, and spin discussed in Refs.~\cite{Lorce:2018zpf,Lorce:2021gxs}.

Similarly, we find that the transverse boost sum rules 
\begin{align}
&   K_{X}^{a}
    (P^{z};s^{\prime},s)
  = \int d^{2}b_{\perp}\,
    K_{X}^{a}
    (\bm{b}_{\perp},P^{z};s^{\prime},s),
\end{align}
are given as
\begin{align}
    K_{E}^{a}
    (P^z;s',s)
& = 0, 
    \label{boost_sum_rule_energy}\\
     K_{M}^{a}
     (P^z;s',s)
& = \frac{\left(\bm{\sigma}_{s^{\prime}s}\times\bm{P}\right)^{a}}{2M},
    \label{boost_sum_rule_mass} \\
    K_{c}^{a}
    (P^{z};s^{\prime},s)
& = \frac{
    \left(
    \bm{\sigma}_{s^{\prime}s}\times\bm{P}
    \right)^{a}
    }{
    2\left(E_{P}+M\right)
    },
    \label{boost_sum_rule_canonical}
\end{align}
where likewise the relations in Eqs.~\eqref{GFFconstraints} and \eqref{spinrelation}, and the normalization of energy distribution in Ref.~\cite{Won:2025dgc} were used.
These results show that the transverse boost sum rules can be understood as the total energy dipole moments defined relative to the relativistic centers, i.e. $\int d^{3}x \,\bigl(R_{X}^{a}-x^{a}\bigr)T^{00}(x)$.

From the above spin and boost sum rules, it is interesting to notice that only $J_{M}^{a}(P^z;s',s)$ and $K_{M}^{a}(P^z;s',s)$ follow the Lorentz transformation rules of an antisymmetric rank-2 tensor~\cite{Lorce:2018zpf}:
\begin{align}
    J_{M}^{a}
& = \frac{E_P}{M}
    {J^a_{M}}_{*}
  + \frac{(\bm{P}\times{\bm{K}_{M}}_{*})^{a}}{M},
    \label{Jlortranrule} \\
    K_{M}^{a}
& = \frac{E_P}{M}
    {K^a_{M}}_{*}
  - \frac{(\bm{P}\times{\bm{J}_{M}}_{*})^{a}}{M},
    \label{Klortranrule}
\end{align}
where $\bm{J}_*$ and $\bm{K}_*$ are the expectation value of TAM and boost in the BF. 
This is expected because only the coordinates of the relativistic center of mass transform as part of a Lorentz four-vector~\cite{Lorce:2021gxs}. Note, however, that these relations do not hold at the level of spatial distributions.

\subsection{Numerical illustration: spatial distribution of transverse TAM and boost}
\label{numerical_discussion_IF}

In this work, we denote the total 2D transverse AM distributions for an unpolarized nucleon in the transverse plane as
\begin{align}
    g
    (\bm{b}_{\perp},P^{z})
&:= \frac{1}{2} 
    \sum_{s^{\prime},s}
    g
    (\bm{b}_{\perp},P^{z};s^{\prime},s)
    \delta_{s^{\prime}s} 
\end{align}
with $g=L^{a},S^{a},J^{a},K^{a}$. 
For a transversely polarized nucleon, we choose spin states along the $x$-axis, i.e., $\ket{s^{x}=\pm\frac{1}{2}}=\left(\ket{s^{z}=+\frac{1}{2}}\pm\ket{s^{z}=-\frac{1}{2}}\right)/\sqrt{2}$, so that the corresponding distributions are given as 
\begin{align}
    g^{T}
(\bm{b}_{\perp},P^{z})
&:= g
    (\bm{b}_{\perp},P^{z};s^{x},s^{x})  .
\end{align}

\begin{figure*}[htbp]
  \centering
  \textbf{2D spatial distributions of transverse TAM of a nucleon}
  
    \vspace{-1.8cm}

  \includegraphics[width=1\textwidth]{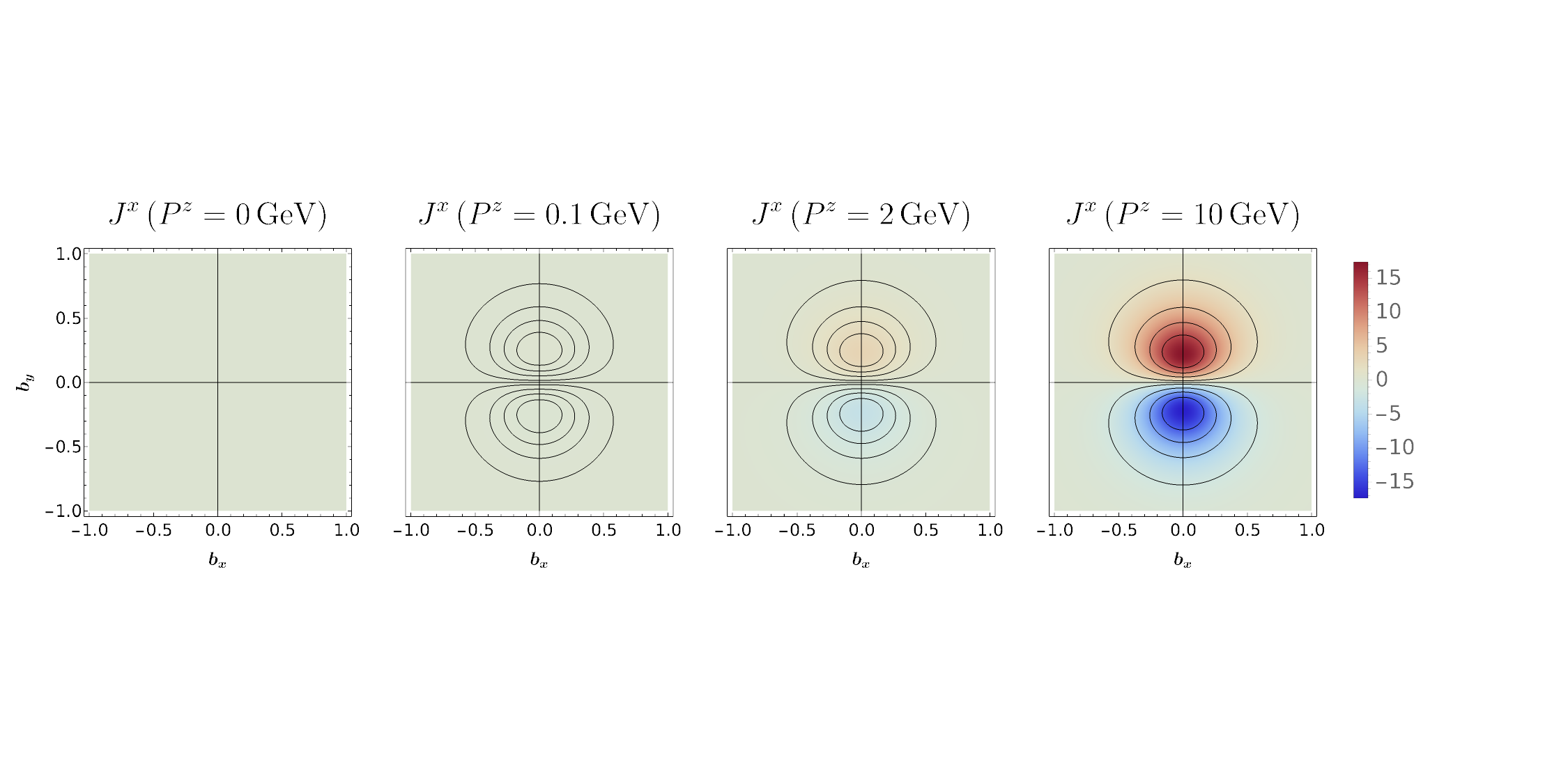}
    \vspace{-4.5cm} 

    \includegraphics[width=1\textwidth]{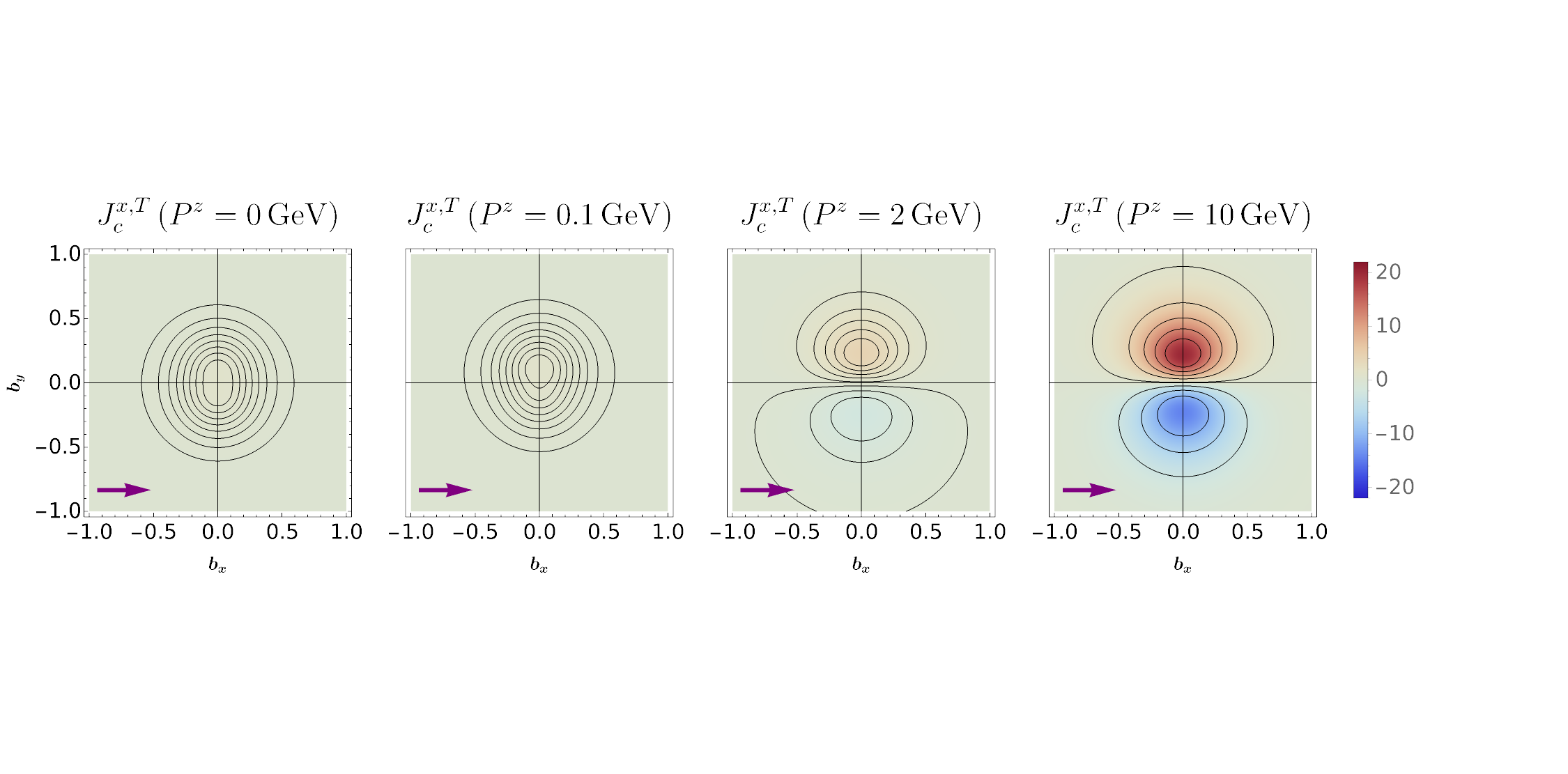}
\vspace{-4.5cm}

    \includegraphics[width=1\textwidth]{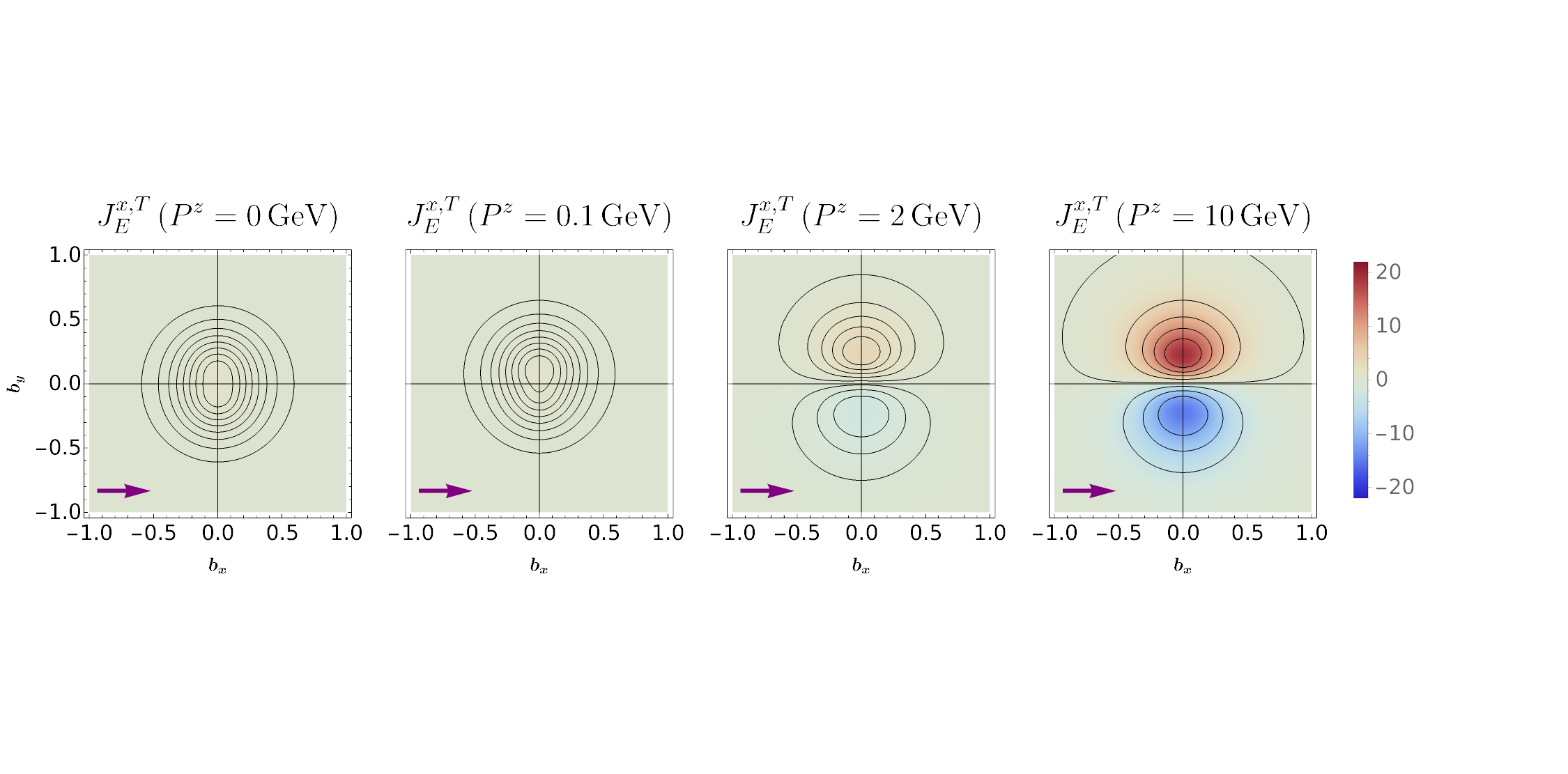}
\vspace{-4.5cm}

  \includegraphics[width=1\textwidth]{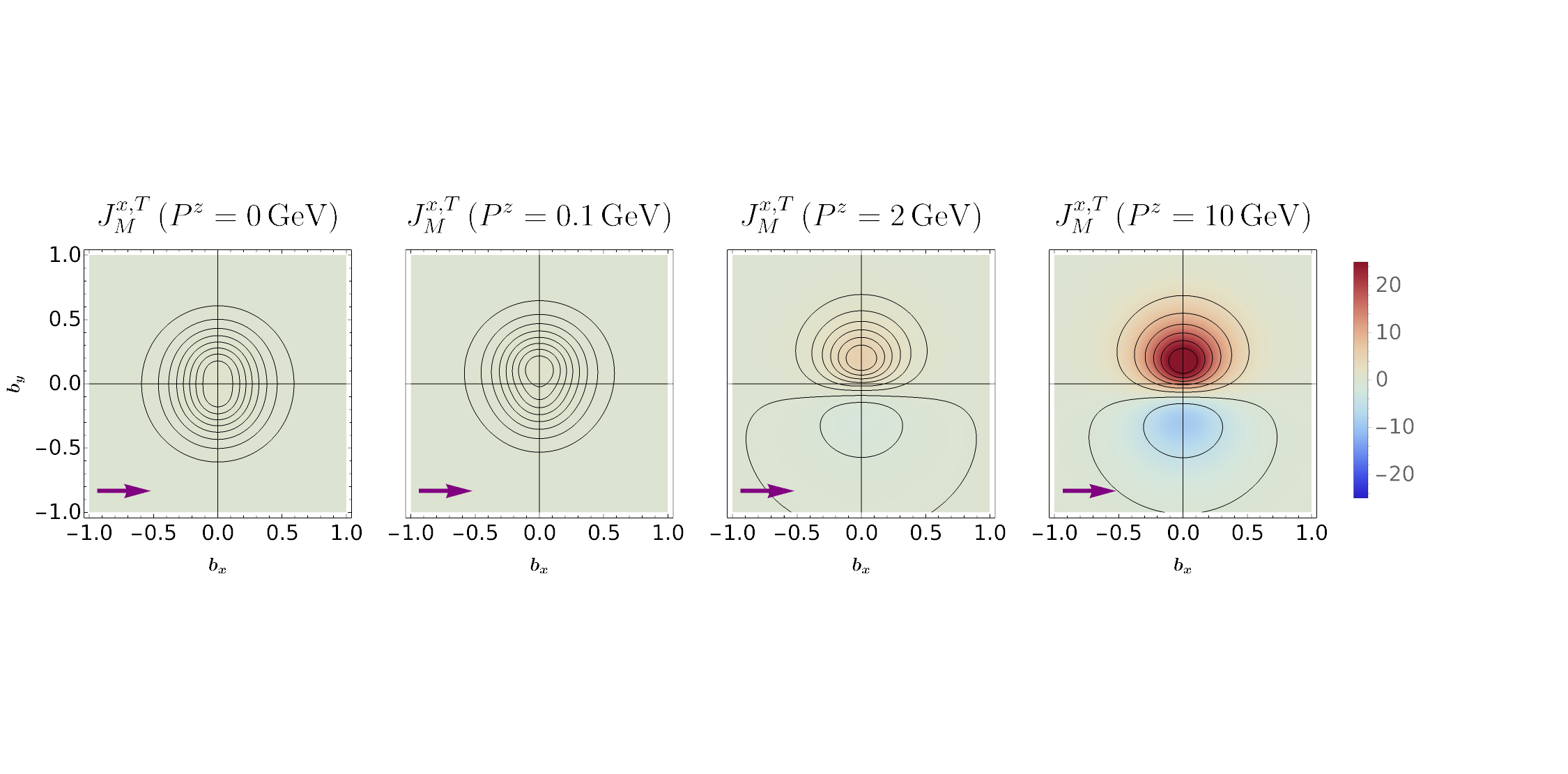}

\vspace{-2cm}

  \caption{Total (i.e.,~quark + gluon) spatial distributions of transverse TAM in the transverse plane for an unpolarized nucleon (upper row) and a transversely polarized nucleon along the $x$-axis (lower three rows) for four different values of the nucleon momentum.
  Based on the simple multipole model of Ref.~\cite{Hackett:2023rif} for the EMT FFs, assuming a target mass of $M = 0.938$ GeV.  }
  \label{fig:1}
\end{figure*}

In Fig.~\ref{fig:1}, we illustrate the total (i.e., quark + gluon) distributions of transverse TAM in the transverse plane for an unpolarized nucleon (upper row) and a transversely polarized nucleon along the $x$-axis (lower three rows) as $P^{z}$ increases (see also Ref.~\cite{Lorce:2025pxt} for the decomposition into OAM and intrinsic spin). 
For the unpolarized nucleon, the spin-dependent monopole and quadrupole contributions vanish, and so does the pivot dependence. The upper-row panels exhibit then a pure dipole pattern over the entire range of $P^{z}$.
As a result, the transverse TAM in an unpolarized nucleon vanishes upon integration over the transverse plane.

By contrast, the lower three rows show the corresponding distributions defined relative to the centers of spin, energy, and mass for a transversely polarized nucleon along the $x$-direction. 
For nonzero $P^{z}$, spin-dependent contributions emerge. 

It is evident from Eq~\eqref{def_internal_am_1} that the differences between the spatial distributions of $J_c^{x,T}$, $J_E^{x,T}$, and $J_M^{x,T}$ in the large-$P^z$ limit (see rightmost panels of bottom three rows of Fig.~\ref{fig:1}) are entirely driven by the pivot-dependent term $-R_X^y P^z(\mathbf{b}_\perp)$. 
Due to the predominantly positive monopole structure of the longitudinal momentum density~\cite{Won:2025dgc}, this contribution distorts the dipole pattern of $J_c^{x,T}$ according to the choice of relativistic center. 
For the center-of-mass pivot (see Eqs.~\eqref{RM} and \eqref{def_internal_am_1}), the pivot contribution grows with boost and increasingly converts the negative part of the dipole into a positive contribution close to the origin, producing the downward displacement of the dipolar pattern observed in $J_M^{x,T}$. 
For the center-of-energy pivot (see Eqs.~\eqref{RE} and \eqref{def_internal_am_1}), the correction has the opposite sign; and additionally, while $R_E^y$ vanishes at large $P^z$, the simultaneous growth of $P^z(\mathbf{b}_\perp)$ leaves a finite residual effect. Consequently, the net transverse angular momentum decreases with increasing hadron momentum, in agreement with the center-of-energy spin sum rule.

\begin{figure*}[htbp]
  \centering
  \textbf{2D spatial distributions of transverse boost of a nucleon}
  
    \vspace{-1.8cm}

   \includegraphics[width=1\textwidth]{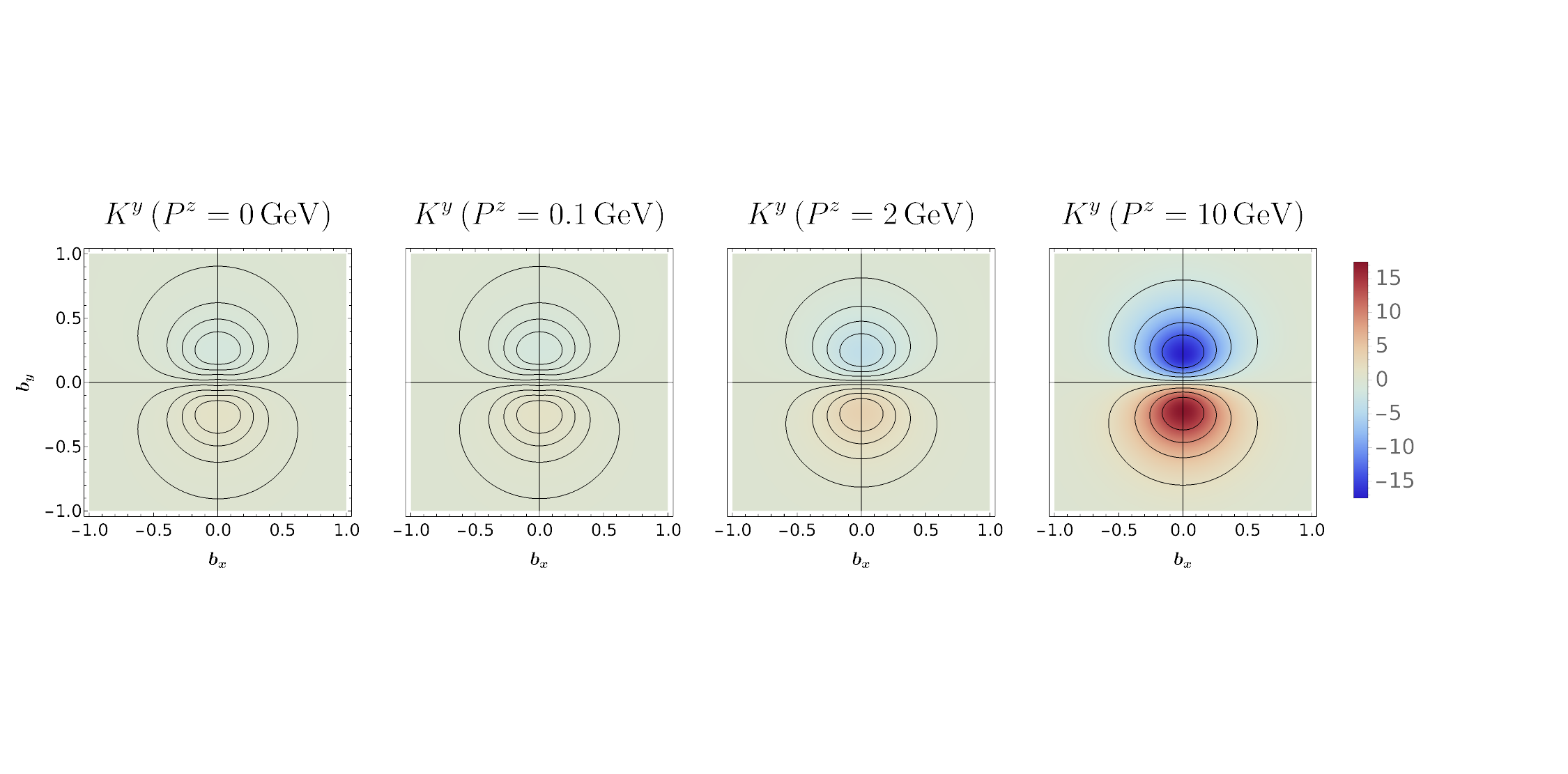}

    \vspace{-4.5cm}
    
   \includegraphics[width=1\textwidth]{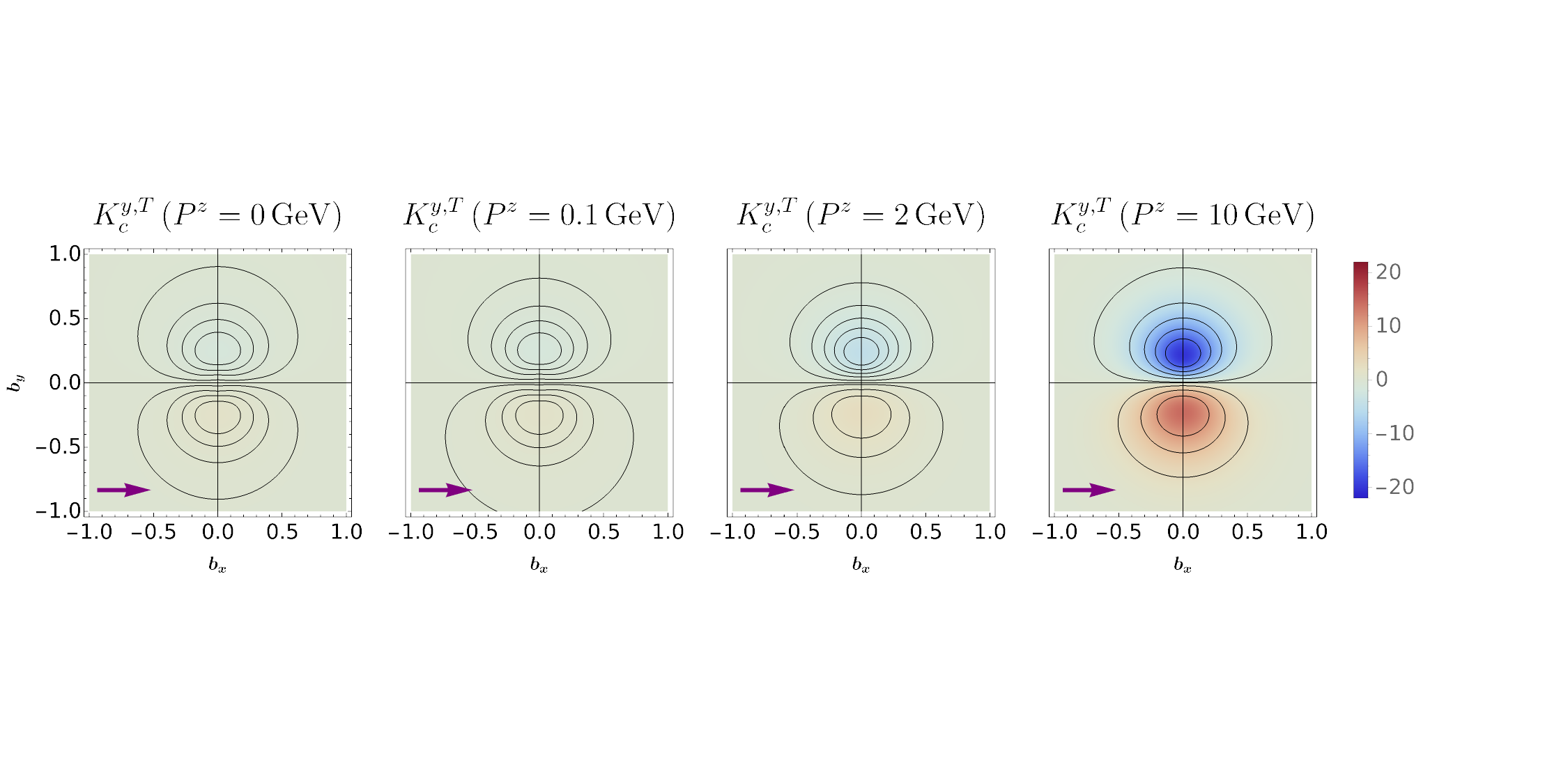}
    \vspace{-4.5cm}

    \includegraphics[width=1\textwidth]{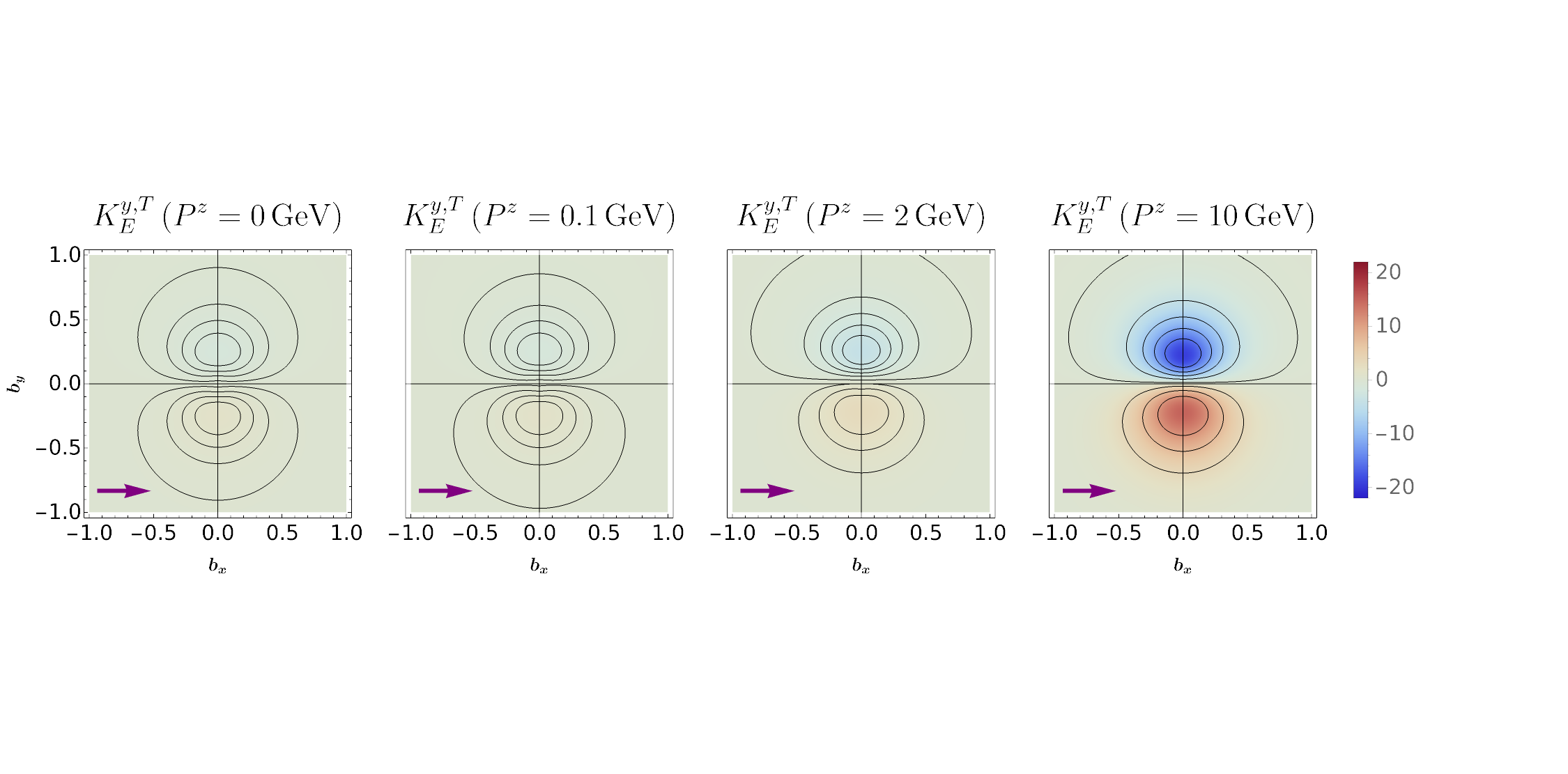}
    \vspace{-4.5cm}

  \includegraphics[width=1\textwidth]{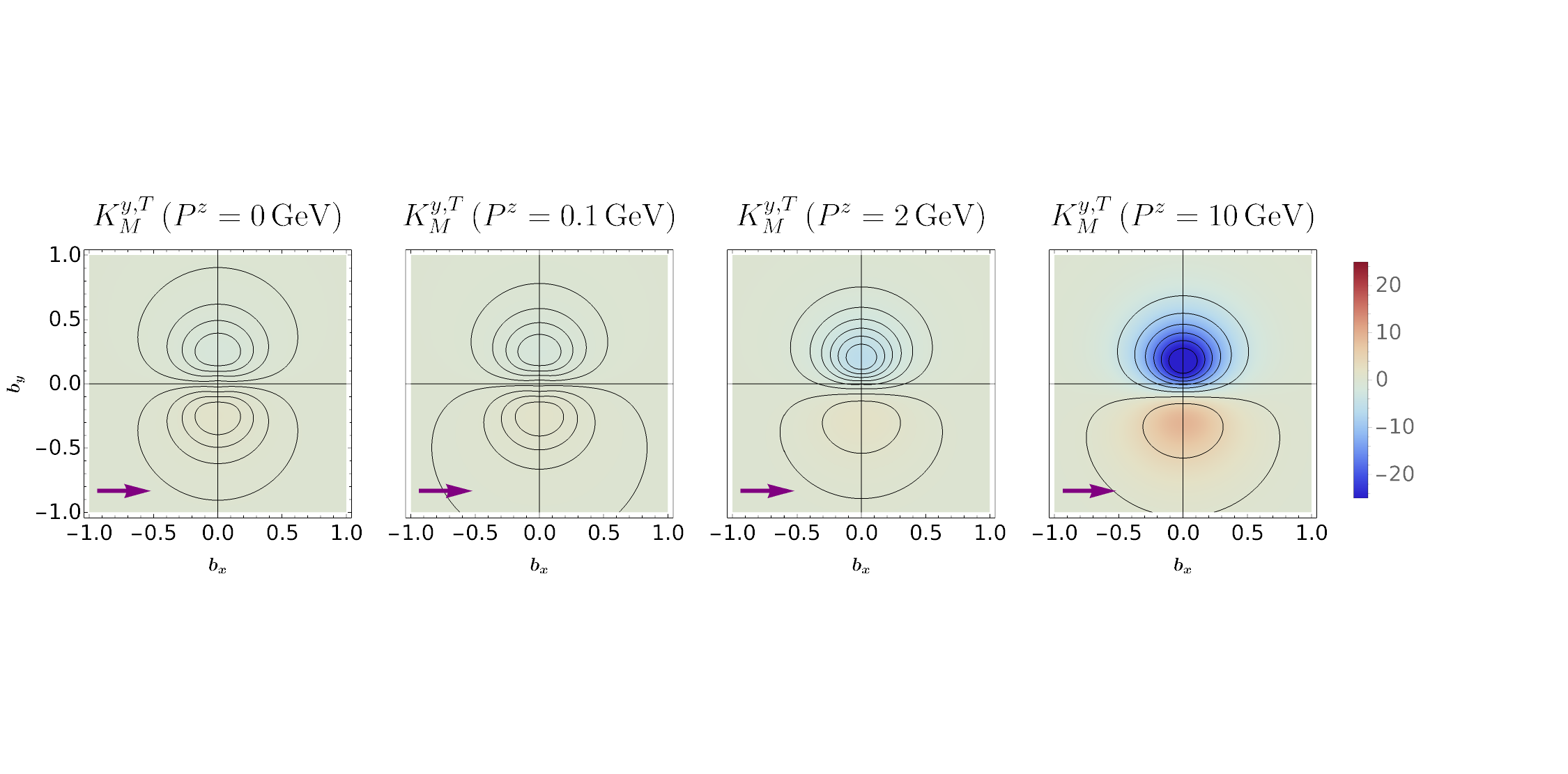}

  \vspace{-2cm}

  \caption{Total (i.e.,~quark + gluon) spatial distributions of transverse boost on the transverse plane for an unpolarized nucleon (upper row) and a transversely polarized nucleon along the $x$-axis (lower three rows) for four different values of the nucleon momentum.
  Based on the simple multipole model of Ref.~\cite{Hackett:2023rif} for the EMT FFs, assuming a target mass of $M = 0.938$ GeV.}
  \label{fig:2}
\end{figure*}

In Fig.~\ref{fig:2}, we show the corresponding transverse boost distributions in the transverse plane, organized in the same way as Fig.~\ref{fig:1}. 
As in the transverse TAM case, the upper-row panels exhibit a pure dipole pattern for the unpolarized nucleon, implying that the distributions vanish upon integration over the transverse plane. For the transversely polarized nucleon along the $x$-direction, however, the lower three rows reveal a slightly different behavior from that observed for the transverse TAM distributions. Specifically, Eq~\eqref{def_internal_am_2} suggests that the differences between the spatial distributions of $K_c^{y,T}$, $K_E^{y,T}$, and $K_M^{y,T}$ in the large-$P^z$ limit (see rightmost plots of bottom three rows of Fig.~\ref{fig:2}) are  driven by the pivot-dependent term $R_X^y E(\mathbf{b}_\perp)$.
The pivot-dependent terms shift the center of the canonical dipole distribution, and because the spatial distribution of $K_c^{y,T}$ and the pivot-dependent contributions both enter with signs opposite to those in the angular-momentum case, the resulting distortions are same. As a consequence, the dipolar patterns of $K_M^{y,T}$ and $K_E^{y,T}$ are again displaced in the same direction as that observed for $J_M^{x,T}$ and $J_E^{x,T}$.
As in the transverse TAM case, the boost sum rules in Eqs.~\eqref{boost_sum_rule_energy} and \eqref{boost_sum_rule_mass} are already manifest at the distribution level.

\begin{figure*}[htbp]
  \centering
  \textbf{2D spatial distributions of $J^y$ and $K^x$ of a nucleon}
  
    \vspace{-1.8cm}

   \includegraphics[width=1\textwidth]{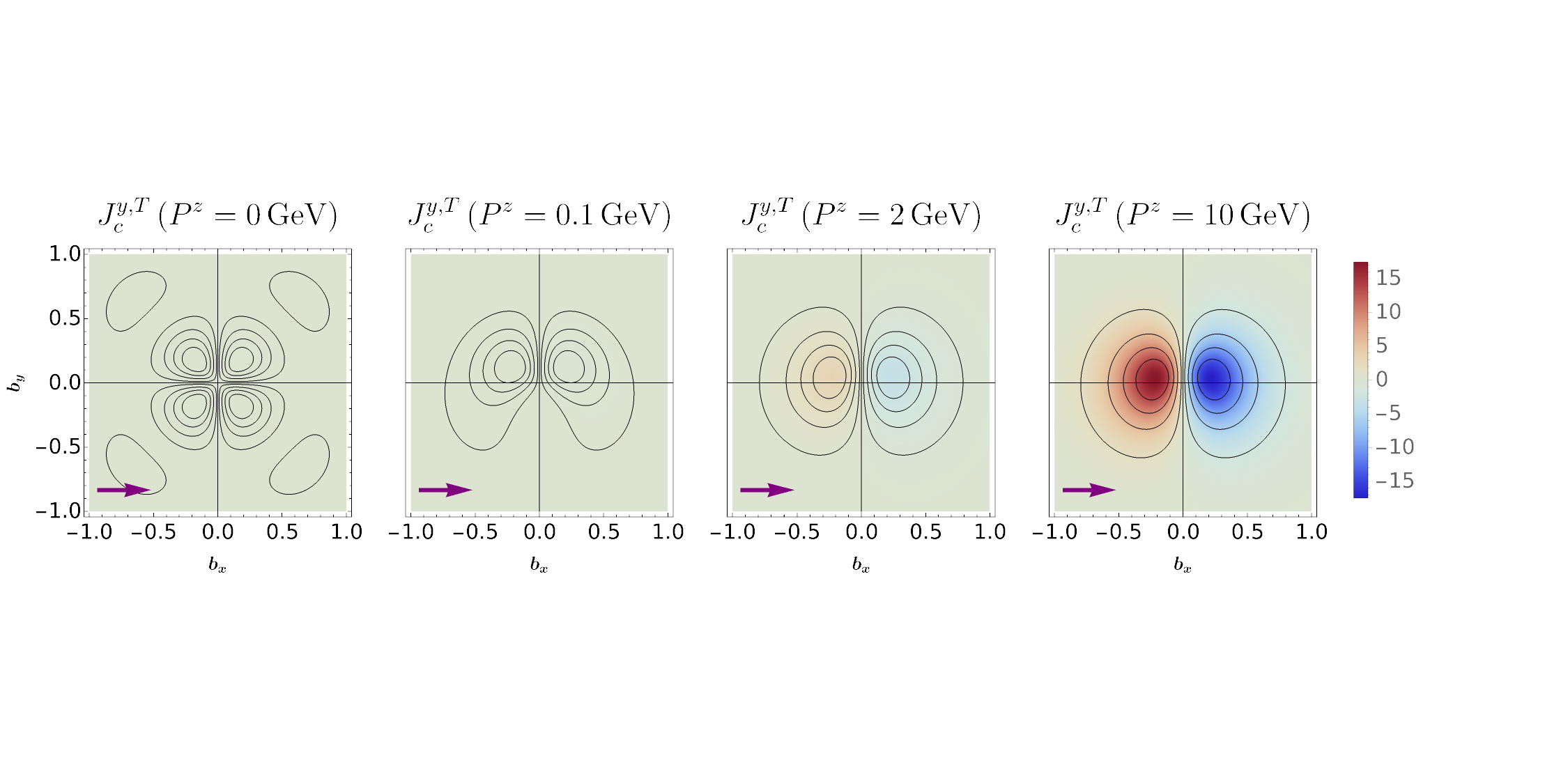}

    \vspace{-4.5cm}
    
   \includegraphics[width=1\textwidth,trim=0cm 1cm 0cm 0cm,clip]{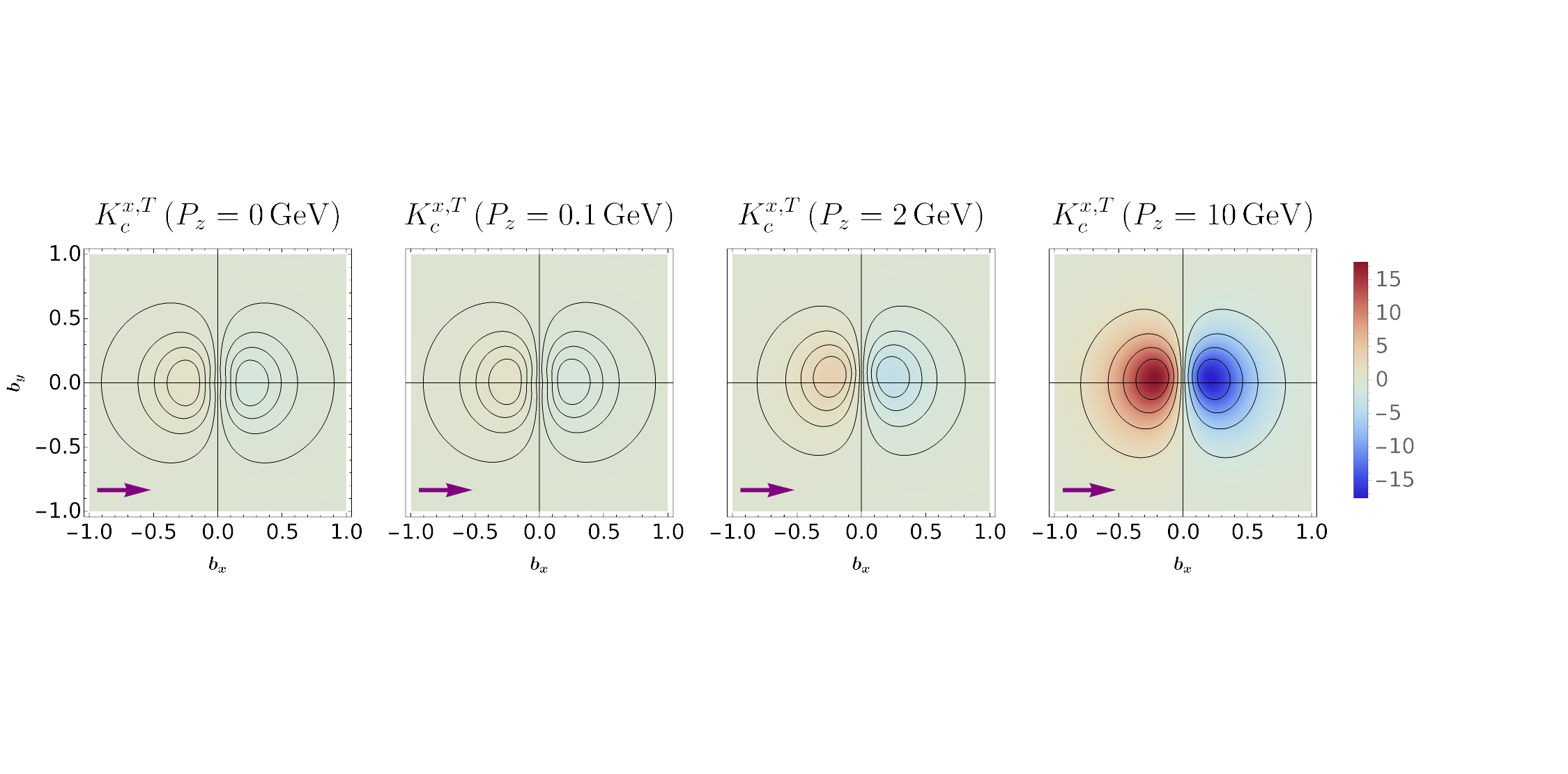}

  \vspace{-2cm}

  \caption{Total (i.e.,~quark + gluon) spatial distributions of $J^y$ (upper row) and $K^x$ (lower row) on the transverse plane for a transversely polarized nucleon along the $x$-axis for four different values of the nucleon momentum.
  Based on the simple multipole model of Ref.~\cite{Hackett:2023rif} for the EMT FFs, assuming a target mass of $M = 0.938$ GeV.}
  \label{fig:3}
\end{figure*}

We finally display the spatial distributions of the remaining transverse components of TAM and boost, namely, $J^{y,T}$ (upper panel) and $K^{x,T}$ (lower panel) in Fig.~\ref{fig:3}. For a nucleon polarized along the $x$-direction, these turn out to be pivot-independent. Nevertheless, the spatial structures exhibited by these components are also interesting. Specifically, there is no monopole contribution to $J^{y,T}$ and $K^{x,T}$, as is also evident from Eqs.~\eqref{Jk_spin1/2} and ~\eqref{Kk_spin1/2}. Thus, $J^{y,T}(\bm{b_\perp})$ is a pure quadrupole in the Breit frame. As the nucleon momentum increases, dipole contributions are introduced and become increasingly dominant in the large $P^z$-limit. In contrast, the boost distribution is a dipole in the Breit frame and stays predominantly the same as we increase $P^z$. The quadrupole contribution only slightly distorts the distribution leading to a small upward shift of the dipole axis in both $J^{y,T} (\bm{b_\perp})$ and $K^{x,T}(\bm{b_\perp})$.

For a spin-$0$ target, all the target spin-dependent features observed in the nucleon case disappear, making the analysis much simpler. Analytical expressions and numerical illustrations of the spatial distributions of TAM and boost inside a pion are presented in Appendix~\ref{app:pion}.

\section{Light-front formalism}
One of the key features of the LF formalism is to preserve Galilean symmetry in the transverse plane, see Refs.~\cite{Susskind:1967rg,Kogut:1969xa,Harindranath:1996hq,Brodsky:1997de}.
Due to this symmetry, spatial distributions free of relativistic recoil corrections can be defined within this framework~\cite{Burkardt:2000za,Burkardt:2002hr}, which can also be interpreted in terms of probability densities.
These have been used to map the electric charge~\cite{Miller:2007uy,Carlson:2007xd} and current~\cite{Chen:2022smg}, polarization and magnetization~\cite{Miller:2007kt,Miller:2010nz,Chen:2023dxp}, longitudinal LF momentum~\cite{Abidin:2008sb}, boost~\cite{Burkardt:2005hp} and AM~\cite{Lorce:2017wkb}, and more generally the EMT~\cite{Lorce:2018egm,Freese:2021czn,Won:2025dgc,Won:2026ljg} inside the nucleon.

In the LF formalism, the LF components of a four-vector are given as 
\begin{align}
    a^\mu = \left( a^+, a^-, \bm{a}_\perp \right),
\end{align}
where $a^\pm=(a^0\pm a^3)/\sqrt{2}$.
The inner product between position and momentum vectors reads 
\begin{align}
    x \cdot p  
& = x^{+} p^{-} 
  + x^{-} p^{+} 
  - \boldsymbol{x}_\perp \cdot \boldsymbol{p}_\perp.
\end{align}
Following this relation, we denote the LF time and energy by $x^+$ and $p^{-}$, and the longitudinal LF coordinate and momentum by $x^-$ and $p^+$, respectively~\cite{Harindranath:1996hq}.

\subsection{Definition of LF generalized AM operators}
The LF generalized AM operators are obtained from the covariant definition
in Eq.~\eqref{def_AM} by choosing the null hypersurface at $x^{+}=\mathrm{constant}$, whose normal vector is $n^{\mu}=(1,0,0,-1)/\sqrt{2}$.
Following the classification of the LF Poincaré generators into
kinematical and dynamical ones, the transverse LF TAM and boost
operators are defined as~\cite{Kogut:1969xa,Lorce:2018zpf}
\begin{align}
    \mathcal{\hat{J}}^{a}
& = \mathcal{\hat{L}}^{a} + \mathcal{\hat{S}}^{a},\\
    \mathcal{\hat{B}}^{a}
& = \int d^{2}x_{\perp}dx^{-}\,
    \hat{J}^{++a}(x),
    \label{def_LF_transverse_boost} 
\end{align}
where the transverse LF OAM $\mathcal{\hat{L}}^{a}$ and intrinsic spin $\mathcal{\hat{S}}^{a}$ are
\begin{align}
    \mathcal{\hat{L}}^{a}
& = \epsilon_{\perp}^{ab}
    \int d^{2}x_{\perp}dx^{-}\,
    \hat{L}^{+-b}(x), 
    \label{def_LF_transverse_OAM}   \\
    \mathcal{\hat{S}}^{a}
& = \epsilon_{\perp}^{ab}
    \int d^{2}x_{\perp}dx^{-}\,
    \hat{S}^{+-b}(x).
\end{align}
These transverse LF TAM operators do not satisfy the $su(2)$ algebra and therefore cannot, in principle, be interpreted as genuine spin operators.
Genuine spin operators within the LF formalism have, however, been derived from the Pauli-Lubański pseudovector~\cite{Harindranath:2001rc,Harindranath:2013goa}.
The corresponding spatial distributions are discussed in Appendix~\ref{app_LF_genuine_spin_operator}.

\subsection{Light-front center of the system}

In the LF formalism, the standard choice for the center of the system is the center of longitudinal LF momentum~\cite{Burkardt:2002hr,Lorce:2018zpf}.
In the symmetric LF frame, $\bm{P}_{\perp}=\bm{0}_{\perp}$, the LF momentum plays the role of inertia in the transverse plane. The corresponding position operator is defined by
\begin{align}
    \hat{R}_{n}^{\mu}
& = \frac{1}{P^{+}}
    \int d^{2}x_{\perp}dx^{-}\,
    x^{\mu}
    \hat{T}^{++} (x),
\end{align}
where the subscript $n$ indicates that the position operator is defined on the hypersurface with fixed value of $x^+=x\cdot n$.
This operator coincides with the IMF limit of the position operator for the canonical center and the relativistic center of energy, 
\begin{align}
    \hat{R}_{n}^{\mu}
  = \lim_{P^{z}\to\infty} 
    \hat{R}_{E}^{\mu}
  = \lim_{P^{z}\to\infty} 
    \hat{R}_{c}^{\mu}.
\end{align}

\subsection{LF spatial distributions}
Analogous to the IF formalism, the derivative with respect to $\Delta^{+}$ appearing in the definition in Eq.~\eqref{def_LF_transverse_OAM} cannot be considered directly in the Drell-Yan frame (DYF) where $\Delta^+=0$ by definition~\cite{Drell:1970wh,West:1970av}. 
We thus define the 2D LF distributions by integrating the corresponding 3D LF distributions derived in the generic LF frame (GLF) over the longitudinal component $x^{-}$.

In the symmetric LF frame, the average momentum and momentum transfer are given by 
\begin{align}
    P^{\mu}
& = (P^{+},P^{-},\bm{0}_{\perp}),   
    \hspace{0.5cm}
    \Delta^{\mu}
  = (\Delta^{+},\Delta^{-},\bm{\Delta}_{\perp}).
\end{align}
The on-shell conditions read 
\begin{align}
    \Delta^{-}
  = - \frac{P^-\Delta^+}{P^+}, \hspace{1cm}
    P^- 
  = \frac{M^2(1+\tau)}{2P^{+}}>0,
    \label{massshell}
\end{align}
where the LF energy transfer $\Delta^{-}$ vanishes only in the DYF.

Following the definition of the expectation value in the quantum phase-space formalism in Eq.~\eqref{def_3D_internal_distribution} with the LF helicity $\lambda^{\prime},\lambda$, the 2D spatial distributions of the transverse LF TAM and boost are given by\footnote{In the LF formalism, the position three-vector corresponds to $\bm{r}=(r^-,\bm{r}_\perp)$ and the momentum three-vector to $\bm{P}=(P^+,\bm{P}_\perp)$.}
\begin{align}
    \hspace{-0.5cm}\mathcal{J}^{a} 
    (\bm{b}_{\perp},P^{+};\lambda^{\prime},\lambda) 
& = \mathcal{L}^{a} 
    (\bm{b}_{\perp},P^{+};\lambda^{\prime},\lambda)  \notag\\
    \hspace{-0.5cm}
& + \mathcal{S}^{a} 
    (\bm{b}_{\perp},P^{+};\lambda^{\prime},\lambda), \\
    \hspace{-0.5cm}\mathcal{B}^{a} 
    (\bm{b}_{\perp},P^{+};\lambda^{\prime},\lambda) 
& =  
  - \int dr^{-}\,
    r^{a} 
    \langle 
    \hat{T}^{++}(\bm{r})
   \rangle_{\bm{0},\bm{P}}^{\lambda^{\prime}\lambda},
\end{align}
where
\begin{align}
    \hspace{-0.5cm}\mathcal{L}^{a} 
    (\bm{b}_{\perp},P^{+};\lambda^{\prime},\lambda) 
& = \epsilon^{ab}_{\perp}  
    \int dr^{-}\,
    r^{-}
    \langle 
    \hat{T}^{+b}(\bm{r})
    \rangle_{\bm{0},\bm{P}}^{\lambda^{\prime}\lambda} \notag\\
    \hspace{-0.5cm}
& - \epsilon^{ab}_{\perp}  
    \int dr^{-}\,
    r^{b}
    \langle 
    \hat{T}^{+-}(\bm{r})
    \rangle_{\bm{0},\bm{P}}^{\lambda^{\prime}\lambda},  \\
    \hspace{-0.5cm}\mathcal{S}^{a} 
    (\bm{b}_{\perp},P^{+};\lambda^{\prime},\lambda) 
& = \epsilon_{\perp}^{ab}
    \int dr^{-}\,
    \langle 
    \hat{S}^{+-b}(\bm{r})
    \rangle_{\bm{0},\bm{P}}^{\lambda^{\prime}\lambda}.
\end{align}
The relative coordinates are defined as $r^{\mu}:=x^{\mu}-R_{n}^{\mu}=(r^{+},\bm{r})$ with $\bm{r}=(r^{-},\bm{b}_{\perp})$, and the distributions are evaluated at fixed LF time $r^{+}=0$.
In the present work, we restrict our analysis to these 2D distributions and do not discuss the corresponding 3D distributions or their physical interpretation.

The transverse LF OAM distribution in the transverse plane is given by
\begin{align}
&   \hspace{-0.5cm}   
    \mathcal{L}^{a}
    (\bm{b}_{\perp},P^{+};\lambda^{\prime},\lambda)
  = \frac{M}{P^{+}}
    \int \frac{d^{2}\Delta_{\perp}}{\left(2\pi\right)^{2}}
    e^{-i\bm{\Delta}_{\perp}\cdot\bm{b}_{\perp}}    \notag\\
&   \hspace{-0.5cm}   \times 
    \left[
    i\delta_{\lambda^{\prime}\lambda}
    \epsilon_{\perp}^{ab}
    \sqrt{\tau}
    X_{1}^{b}
    \tilde{\mathcal{L}}_{1}
  + \sigma_{\lambda^{\prime}\lambda}^{a}
    \tilde{\mathcal{L}}_{0}
  + \sigma_{\lambda^{\prime}\lambda}^{b}
    \tau
    X_{2}^{ba}
    \tilde{\mathcal{L}}_{2}
    \right]
 \label{JLF_spin1/2_1}
\end{align}
with a similar expression for the transverse LF intrinsic spin distribution.
These multipole structures coincide with those of the IF formalism.
The corresponding multipole amplitudes are given by
\begin{align}
    \tilde{\mathcal{L}}_{0}
& = - 2M^{2}
    \frac{d}{dt}
    \left[
    \frac{P^{2}}{2M^{2}}
    \tau
    \left(
    A
  - 2J
    \right)
  + \tau
    \left(
    L
  - F
    \right) 
    \right]
    \notag  \\
& + \frac{\tau}{4}  
    D 
  + \frac{1}{2}
    L, \\ 
    \tilde{\mathcal{L}}_{1}
& = - 4M^{2}
    \frac{d}{dt}
    \left(
    \frac{P^{2}}{2M^{2}}
    A 
  - \tau L
  - F
    \right)
    \notag  \\
& + \frac{1}{2}
    D 
  - \frac{1}{2}
    L, \\
    \tilde{\mathcal{L}}_{2}
& = 4M^{2}
    \frac{d}{dt}
    \left[
    \frac{P^{2}}{2M^{2}}
    \left(
    A
  - 2J
    \right) 
  + L
  - F
    \right] 
     - \frac{1}{2}
    D, \\
    \tilde{\mathcal{S}}_{0}
& = 
    \frac{1}{2}
    G_{A}
  - \frac{\tau}{4}
    G_{P},   \\
    \tilde{\mathcal{S}}_{1}
& = 
  - \frac{1}{2}
    G_{A},   \\    
    \tilde{\mathcal{S}}_{2}
& = 
  - \frac{1}{2}
    G_{P},
\end{align}
where $L=J-S$. 

Similarly, the transverse LF boost distribution is obtained in the transverse plane via
\begin{align}
&   \hspace{-0.6cm}
    \mathcal{B}^{a}
    \left(\bm{b}_{\perp},P^{+};\lambda^{\prime},\lambda\right)
  = \frac{P^{+}}{M}
    \int \frac{d^{2}\Delta_{\perp}}{\left(2\pi\right)^{2}}
    e^{-i\bm{\Delta}_{\perp}\cdot\bm{b}_{\perp}}    \notag\\
&   \hspace{-0.6cm}\times 
    \Big[
   i\delta_{\lambda^{\prime}\lambda}
    \sqrt{\tau}
    X_{1}^{a}
    \tilde{\mathcal{B}}_{1} 
  + \epsilon_{\perp}^{ab}
    \sigma_{\lambda^{\prime}\lambda}^{b}
    \tilde{\mathcal{B}}_{0}
  + \epsilon_{\perp}^{bc}
    \sigma_{\lambda^{\prime}\lambda}^{b}
    \tau
    X_{2}^{ca}
    \tilde{\mathcal{B}}_{2}
    \Big],
\label{BLF_spin1/2_1}
\end{align}
where
\begin{align}
    \tilde{\mathcal{B}}_{0}
& = 2M^{2}
    \frac{d}{dt}
    \left[
    \tau
    \left(
    A
  - 2
    J
    \right)
    \right], \\ 
    \tilde{\mathcal{B}}_{1}
& = - 4M^{2}
    \frac{dA}{dt}, \\
    \tilde{\mathcal{B}}_{2}
& = - 4M^{2}
    \frac{d}{dt}
    \left(
    A
  - 2
    J
    \right).
\end{align}
These multipole amplitudes involve only $A$ and $J$ FFs.

\subsection{Sum rules: transverse LF spin and boost}

Integrating the above distributions over the transverse plane, we obtain
\begin{align}
    \mathcal{L}^{a}
    (P^{+};\lambda^{\prime},\lambda)
& = \int d^{2}b_{\perp}\,
    \mathcal{L}^{a}
    (\bm{b}_{\perp},P^{+};\lambda^{\prime},\lambda) \notag\\
& = \frac{M}{P^{+}}
\frac{\sigma_{\lambda^{\prime}\lambda}^{a}}{2}
    \left[
    1
  - G_A(0)
    \right], \\
    \mathcal{S}^{a}
    (P^{+};\lambda^{\prime},\lambda)
& = \int d^{2}b_{\perp}\,
    \mathcal{S}^{a}
    (\bm{b}_{\perp},P^{+};\lambda^{\prime},\lambda)\notag\\
& = \frac{M}{P^{+}}
    \frac{\sigma_{\lambda^{\prime}\lambda}^{a}}{2}
    G_{A}(0),
\end{align}
where we have used the FF relations in Eqs.~\eqref{GFFconstraints} and \eqref{spinrelation}.
We thus reproduce the transverse LF spin sum rule~\cite{Lorce:2018zpf}
\begin{align}
    \mathcal{J}^{a}
    (P^{+};\lambda^{\prime},\lambda)
& = \mathcal{L}^{a}
    (P^{+};\lambda^{\prime},\lambda)
  + \mathcal{S}^{a}
    (P^{+};\lambda^{\prime},\lambda)\notag\\
& = \frac{M}{P^{+}}
\frac{\sigma_{\lambda^{\prime}\lambda}^{a}}{2}.
    \label{integral_LF_TAM}
\end{align}
This result has the same large-momentum behavior as the transverse spin sum rule relative to the center of energy in Eq.~\eqref{spin_sum_rule_energy} due to the similarity between the LF center and the center of energy~\cite{Lorce:2018zpf}.

The transverse LF boost sum rule~\cite{Brodsky:2000ii} is obtained by integrating the corresponding distribution over the transverse plane
\begin{align}
    \mathcal{B}^{a}
    (P^{+};\lambda^{\prime},\lambda)
& = \int d^{2}b_{\perp}\,
    \mathcal{B}^{a}
    (\bm{b}_{\perp},P^{+};\lambda^{\prime},\lambda)\notag\\
& = - \frac{P^{+}}{M}
    \epsilon_{\perp}^{ab}
    \frac{\sigma_{\lambda^{\prime}\lambda}^{b}}{2}
    \left[
    A(0)
  - 2J(0)
    \right] \notag\\
& = 0. 
  \label{integral_LF_boost}
\end{align}
Likewise, the transverse LF boost sum rule is consistent with the boost sum rule relative to the center of energy in Eq.~\eqref{boost_sum_rule_energy}.

\subsection{Numerical illustration: spatial distribution of transverse LF AM }
\label{numerical_discussion_LF}

We adopt the polarization conventions of Sec.~\ref{numerical_discussion_IF}, with the unpolarized distributions obtained from the trace over LF helicities and the transversely polarized distributions evaluated for the $x$-polarized spin state, i.e., $\ket{s^{x}=\pm 1/2}=   \bigl(    \ket{\lambda=+1/2}    \pm    \ket{\lambda=-1/2}  \bigr)/\sqrt{2}$. 
Therefore, the corresponding distributions are given by
\begin{align}
    h^{T}
    (\bm{b}_{\perp},P^{+})
 := h
    (\bm{b}_{\perp},P^{+};s^{x},s^{x}),
\end{align}
with $h=\mathcal{L},\mathcal{S},\mathcal{J},\mathcal{B}$.

\begin{figure*}[htbp]
        \centering

  \textbf{2D spatial distributions of transverse LF OAM, intrinsic spin, TAM and boost of a nucleon}

    \vspace{-1.2cm}

          \includegraphics[width=1\textwidth,trim=0cm 1.5cm 0cm 1cm,clip]{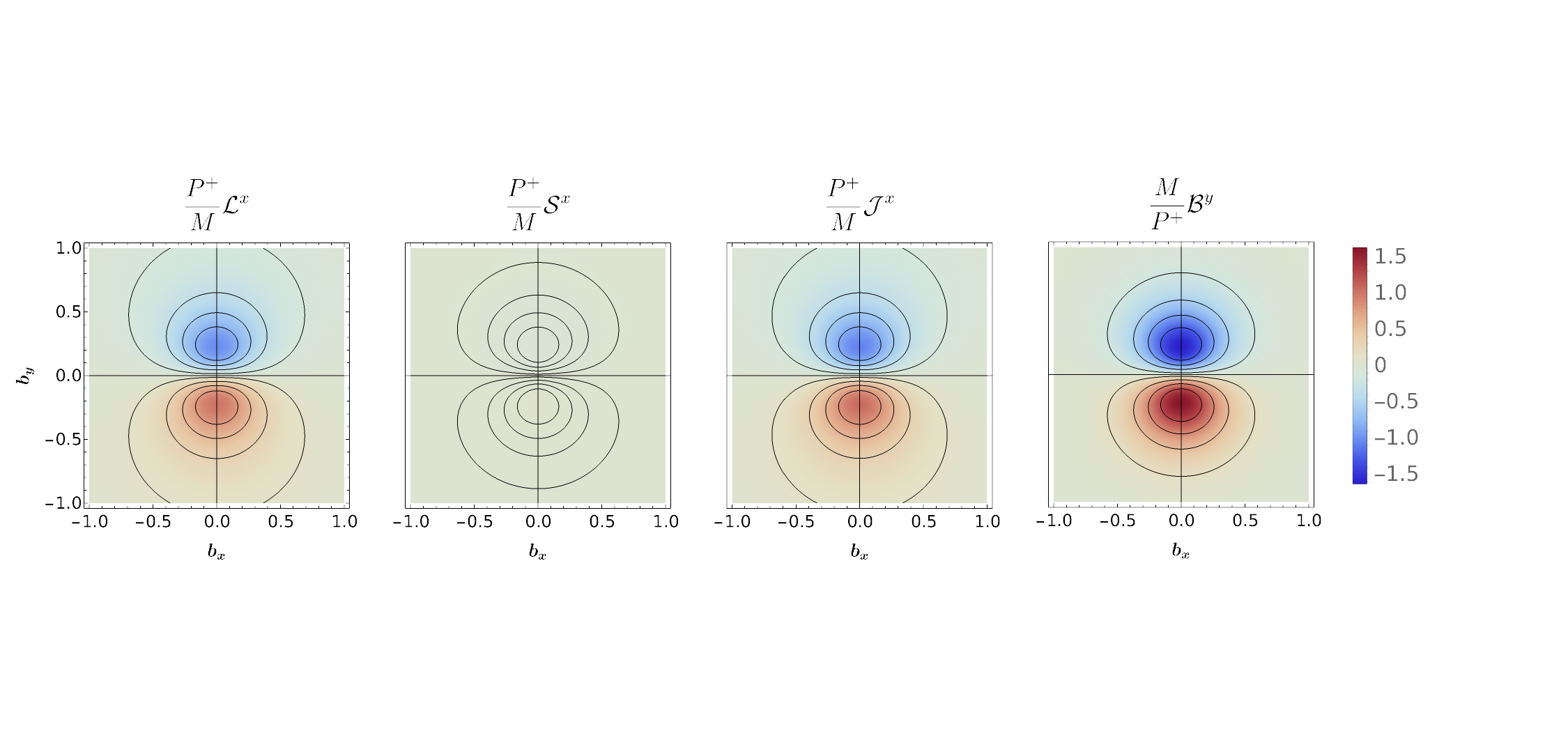}
          \hspace{0cm}
          
          \vspace{-3.5cm} 
          
          \includegraphics[width=1\textwidth,trim=0cm 1.5cm 0cm 0.2cm,clip]{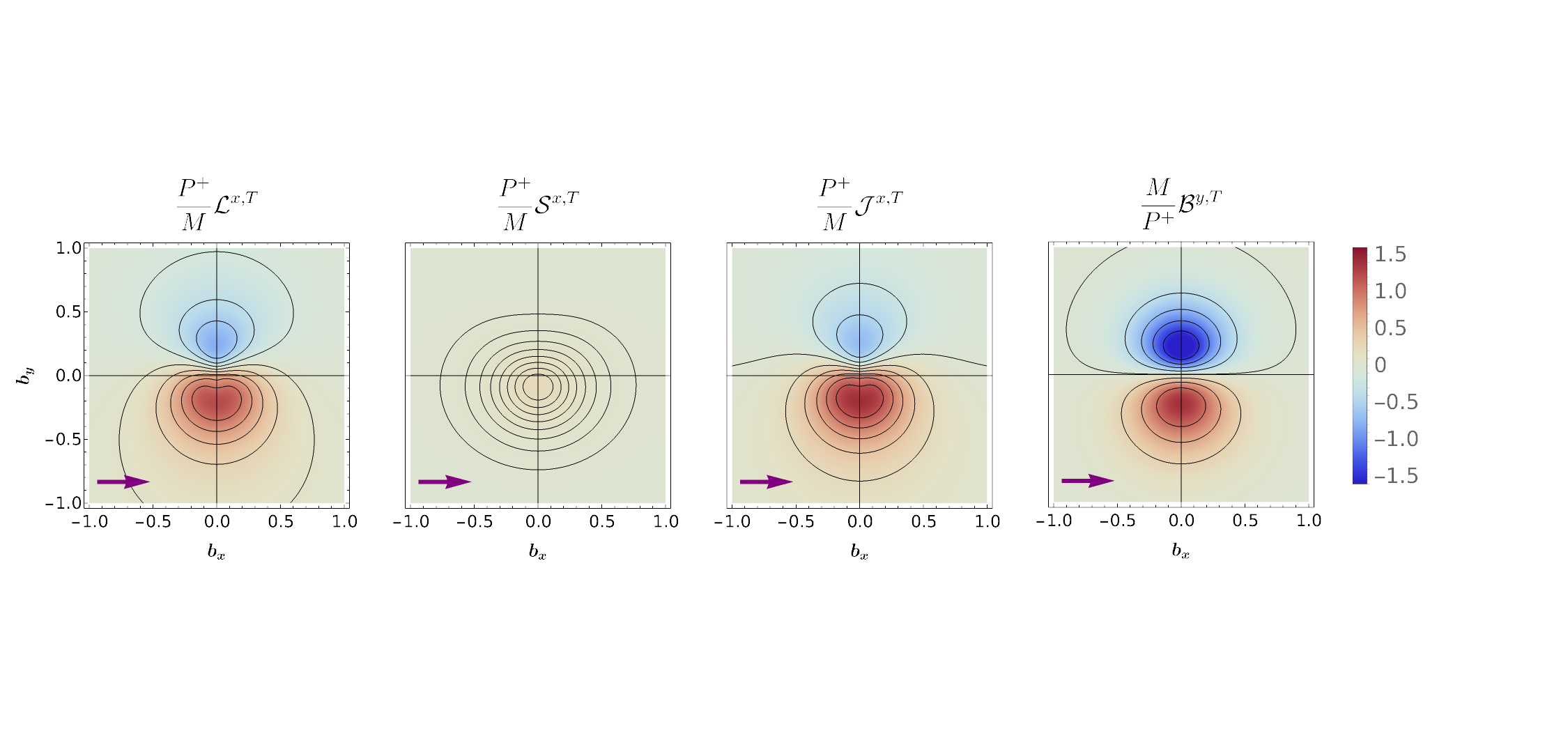}

          \vspace{-1.5cm}

		\caption{Total (i.e.,~quark + gluon) spatial distributions of transverse LF OAM, intrinsic spin, TAM, and boost in the transverse plane for an unpolarized nucleon (upper row) and a transversely polarized nucleon along the $x$-axis (lower row) for four different values of the nucleon momentum.
        A longitudinal LF boost factor has been included to make these distributions $P^+$-independent.
        Based on the simple multipole model of Ref.~\cite{Hackett:2023nkr} for the EMT FFs, assuming a target mass of $M = 0.938$ GeV.}
		\label{fig:4}
\end{figure*}

In Fig.~\ref{fig:4}, we illustrate the spatial distributions of transverse LF OAM, intrinsic spin, TAM, and boost in the transverse plane for a unpolarized nucleon (upper panel) and a transversely polarized nucleon (lower panel) along the $x$-axis in the transverse plane. These distributions are multiplied by an appropriate longitudinal LF boost factor so as to make them $P^{+}$-independent. Comparing the unpolarized case with the transversely polarized case, we observe that the latter exhibit sizable distortions in the transverse plane. 
These distortions arise because the monopole and quadrupole contributions are non-negligible, even though the dipole contribution remains dominant.
We observe that the transverse LF TAM is mainly carried by the transverse LF OAM, analogous to the distributions in the IF formalism for large $P^z$. 

\subsection{Interpolation between the IF and LF formalisms}
The transverse LF boost operator is derived from ``good’’ components of its operator density.
At the level of spatial distributions, a relation between the transverse IF and LF distributions is established in the IMF limit
\begin{align}
    \mathsf{B}^{a}  
    (\bm{b}_{\perp};\lambda^{\prime},\lambda) 
& = \frac{M}{P^+}
    \mathcal{B}^{a}
    (\bm{b}_{\perp},P^{+};\lambda^{\prime},\lambda)  \notag\\
& = \lim_{P^z\rightarrow \infty} 
    \frac{M}{P^z}
    K_{E/c}^{a}
    (\bm{b}_{\perp},P^z;s^{\prime},s)
    \label{boost_EFtoLF},
\end{align}
where $\mathsf{B}^{a}$ is a scaled distribution independent of the longitudinal LF momentum $P^{+}$.
Note that the distinction between the IF and LF spin polarizations disappears in the IMF as a result of the equality between the Melosh and IMF Wigner rotations~\cite{Chen:2022smg}.

By contrast, since the transverse LF TAM involves a ``bad’’ component, one cannot directly compare the transverse IF and LF TAM distributions in the IMF limit.
Instead, we construct the following matching relation in the IMF limit
\begin{align}
    \mathsf{J}^a  
    (\bm{b}_{\perp};\lambda^{\prime},\lambda)
& = \frac{P^{+}}{M} 
    \mathcal{J}^{a}
    (\bm{b}_{\perp},P^{+};\lambda^{\prime},\lambda)  \notag  \\
& = \lim_{P^{z}\to\infty} 
    \frac{P^z}{M}
    \Bigl[
    J_{E/c}^{a}
    (\bm{b}_{\perp},P^{z};s^{\prime},s) \notag \\
& + \epsilon_{\perp}^{ab}
    K_{E/c}^{b}
    (\bm{b}_{\perp},P^{z};s^{\prime},s)
    \Bigr].
    \label{TAM_BFtoIMF}
\end{align}
These relations are only valid when the right-hand side is defined relative to the center of energy or spin.
Expanding the right-hand side in the IMF limit, one finds that the multipole amplitudes inside the brackets all start at order $1/P^{z}$ since the leading terms cancel in this combination.

\section{Summary and Conclusions}

In this work, we have investigated the spatial distributions of transverse generalized angular momentum, including total angular momentum (i.e., orbital angular momentum+intrinsic spin), and boost for spin-$1/2$ targets. 
Using the quantum phase-space formalism, we constructed 2D distributions in the transverse plane by projecting the generic-frame distributions onto the transverse plane. 
We derived the transverse orbital angular momentum, intrinsic spin, total angular momentum, and boost distributions in the instant-form formalism, and extended the analysis to different choices of pivot, namely the centers of energy, mass, and spin. 
For the nucleon, we verified the corresponding transverse spin and boost sum rules for these pivots and showed how the pivot-dependent terms are generated by the spatial distributions of energy and longitudinal momentum.

We also formulated the corresponding transverse AM and boost distributions in the light-front formalism. 
With an appropriate longitudinal light-front boost factor, these light-front distributions become independent of longitudinal light-front momentum. 
We established their connection with the instant-form distributions in the infinite momentum frame limit: the scaled transverse light-front boost is recovered directly from the scaled instant-form boost distribution, whereas the transverse light-front total angular momentum is consistent with an appropriate combination of the light-front total angular momentum and boost distributions. 
This provides a consistent relation between the instant-form and light-front descriptions of transverse angular momentum.

\section{Acknowledgments}
This work is supported by the France Excellence scholarship 
through Campus France funded by the French government 
(Ministère de l’Europe et des Aﬀaires Étrangères), 
Grant No. 141295X (H.-Y.W.). 
All authors also thank the SCPP project RD/0523-IOE00I0-086 from IRCC, IIT Bombay for funding.

\appendix

\section{External AM in the quantum phase-space formalism}
\label{app:external_TAM}

Defining the external TAM operator
\begin{align}
    \hat{J}_{\mathrm{ext},X}^{i}
  = \epsilon^{ijk} \hat{R}_{X}^{j} \hat{P}^{k}
  = \frac{1}{2}
    \epsilon^{ijk} \{\hat{R}_{X}^{j},\hat{P}^{k}\} ,
\end{align}
one can obtain its representation in the quantum phase-space formalism
\begin{align}
&   \hspace{-0.5cm}
    \langle \hat{J}_{\mathrm{ext},X}^{i} \rangle_{\bm{\mathcal{R}},\bm{P}}^{s^{\prime}s}\notag\\
&   \hspace{-0.5cm}
  = \frac{1}{2}
    \epsilon^{ijk}
    \int \frac{d^{3}\Delta}{(2\pi)^{3}}\,
    e^{i\bm{\Delta}\cdot\bm{\mathcal{R}}}
    \frac{\mel{p^{\prime},s^{\prime}}{\{\hat{R}_{X}^{j},\hat{P}^{k}\}}{p,s}}
    {2\sqrt{p^{\prime0}p^{0}}}.
\end{align}
Inserting the projector onto the on-shell one-nucleon subspace,
\begin{align}
    \hspace{-0.5cm}
    \sum_{s^{\prime\prime}}
    \int \frac{d^{3}k}{(2\pi)^{3}}\,
    \frac{\ketbra{k,s^{\prime\prime}}{k,s^{\prime\prime}}}{2k^{0}},
    \quad
    k^{0}=\sqrt{M^{2}+|\bm{k}|^{2}},
    \label{one-nucleon}
\end{align}
changing the integration variables to $\Delta_{1}:=p^{\prime}-k$ and $\Delta_{2}:=k-p$, and using the definition of the four-momentum operator, $\hat{P}^{\mu}  = \int d^{3}x\,\hat{T}^{0\mu}(x)$, one obtains 
\begin{align}
    \langle \hat{J}_{\mathrm{ext},X}^{i} \rangle_{\bm{\mathcal{R}},\bm{P}}^{s^{\prime}s}
& = \int d^{3}x\,
    \langle \hat{J}_{\mathrm{ext},X}^{i}(\bm{x}) \rangle_{\bm{\mathcal{R}},\bm{P}}^{s^{\prime}s}
\end{align}
with the corresponding spatial distribution of the external TAM is given by
\begin{align}
&   \langle \hat{J}_{\mathrm{ext},X}^{i}(\bm{x}) \rangle_{\bm{\mathcal{R}},\bm{P}}^{s^{\prime}s}\notag\\
&:= \frac{1}{2}
    \epsilon^{ijk}
    \sum_{s^{\prime\prime}}
    R_{X}^{j}(s^{\prime},s^{\prime\prime})
    P^{k}(\bm{x}-\bm{\mathcal{R}};s^{\prime\prime},s)   \notag\\
& + \frac{1}{2}
    \epsilon^{ijk}
    \sum_{s^{\prime\prime}}
    P^{k}(\bm{x}-\bm{\mathcal{R}};s^{\prime},s^{\prime\prime})
    R_{X}^{j}(s^{\prime\prime},s).
\end{align}
Although the insertion in Eq.~\eqref{one-nucleon} is not the identity operator in the full Hilbert space, it is exact here because the action of $\hat P^\mu$ on the external one-nucleon momentum eigenstates restricts the intermediate-state sum to the one-nucleon sector. The expressions for $R_{X}^{j}$ are found in Eqs.~\eqref{RE}-\eqref{Rc}, and $P^{k}(\bm{x}-\bm{\mathcal{R}};s^{\prime\prime},s)$ is the momentum distribution (see Ref.~\cite{Won:2026ljg} for the transverse components and Ref.~\cite{Won:2025dgc} for the longitudinal one).
For the external boost, the same procedure is applied and its spatial distribution is derived along the same lines.

\section{Transverse AM and boosts distributions for a pion in the transverse plane} 
\label{app:pion}
The matrix elements of the EMT for a spin-0 target can be parametrized as~\cite{Pagels:1966zza,Hudson:2017xug}
\begin{align}
    \mel{p^{\prime}}{\hat{T}_{a}^{\mu \nu}}{p}
& = 2P^{\mu}P^{\nu}
    A^a
  + \frac{\Delta^{\mu}\Delta^{\nu}
  - g^{\mu\nu}\Delta^{2}}{2}  
    D^a\notag\\
& + 2M^{2} g^{\mu\nu} 
    \bar{C}^a,
\label{EMT_spin0}
\end{align}
where the normalization of the four-momentum states is the same as for a spin‑1/2 target without spin polarizations. To obtain realistic spatial distributions that satisfy momentum and energy sum rules, we adopt a simple multipole model with the tripole ansatz for all EMT FFs based on Ref.~\cite{Hackett:2023nkr} for the pion.\footnote{For the pion, the tripole mass $\Lambda_{F_a}$ is obtained by multiplying the monopole mass by $\sqrt{3}$ so that $dF_a(t)/dt|_{t=0}$ remains unchanged.}

\subsubsection{IF TAM and boost distributions for the pion}
For the pion there are no Wigner rotation effects, so only component mixing arises from Lorentz boosts.
Moreover, since the pion carries no intrinsic spin from quarks and gluons, the TAM distribution coincides with the OAM contribution~\cite{Lorce:2025pxt}, and both the TAM and boost distributions are independent of the choice of pivot according to Eqs.~\eqref{RE}, \eqref{RM}, and \eqref{Rc}. 
The transverse TAM and boost distributions using the spin-0 EMT parametrization given in Eq.~\eqref{EMT_spin0} read
\begin{align}
    \hspace{-0.3cm}J^{a}
    (\bm{b}_{\perp},P^{z})  
 = &\int \frac{d^{2}\Delta_{\perp}}{\left(2\pi\right)^{2}}
    e^{-i\bm{\Delta}_{\perp}\cdot\bm{b}_{\perp}}
    i \epsilon_{\perp}^{ab}
    \sqrt{\tau} X_{1}^{b} \notag\\
&   \times 
    \frac{P^{z}}{M}
    \left[
    4 M^2
    \frac{dA}{dt}
  + \frac{M^{2}}{2\left(P^{0}\right)^{2}}
    D
    \right],
    \label{Lk_spin0}\\
    \hspace{-0.3cm}K^{a}
    (\bm{b}_{\perp},P^{z})
 =&\int \frac{d^{2}\Delta_{\perp}}{\left(2\pi\right)^{2}}
    e^{-i\bm{\Delta}_{\perp}\cdot\bm{b}_{\perp}}
    i \sqrt{\tau} X_{1}^{a} \notag\\
&   \times
    4M^{2}
    \frac{d}{dt}
    \left[
  - \frac{P^{0}}{M}
    A
  + \frac{M}{P^{0}}
    F
    \right].
\label{Kk_spin0}
\end{align}
As $P^{z}$ increases, the first term in Eq.~\eqref{Lk_spin0}, proportional to $dA(t)/dt$, becomes dominant, whereas the second term, proportional to $D(t)$, is progressively suppressed and vanishes in the IMF limit.
The physical interpretation of both the terms is different and is discussed in Ref.~\cite{Lorce:2025pxt} in detail.
By contrast, the transverse boost distribution~\eqref{Kk_spin0} purely arises from the contribution $r^{a}T^{00}$.

\begin{figure*}[htbp]
  \centering\textbf{2D spatial distributions of transverse TAM and boost of a pion}

  \vspace{0cm} 

  \includegraphics[width=1\textwidth,
                 trim=0cm 4.5cm 0cm 3.5cm,
                 clip]{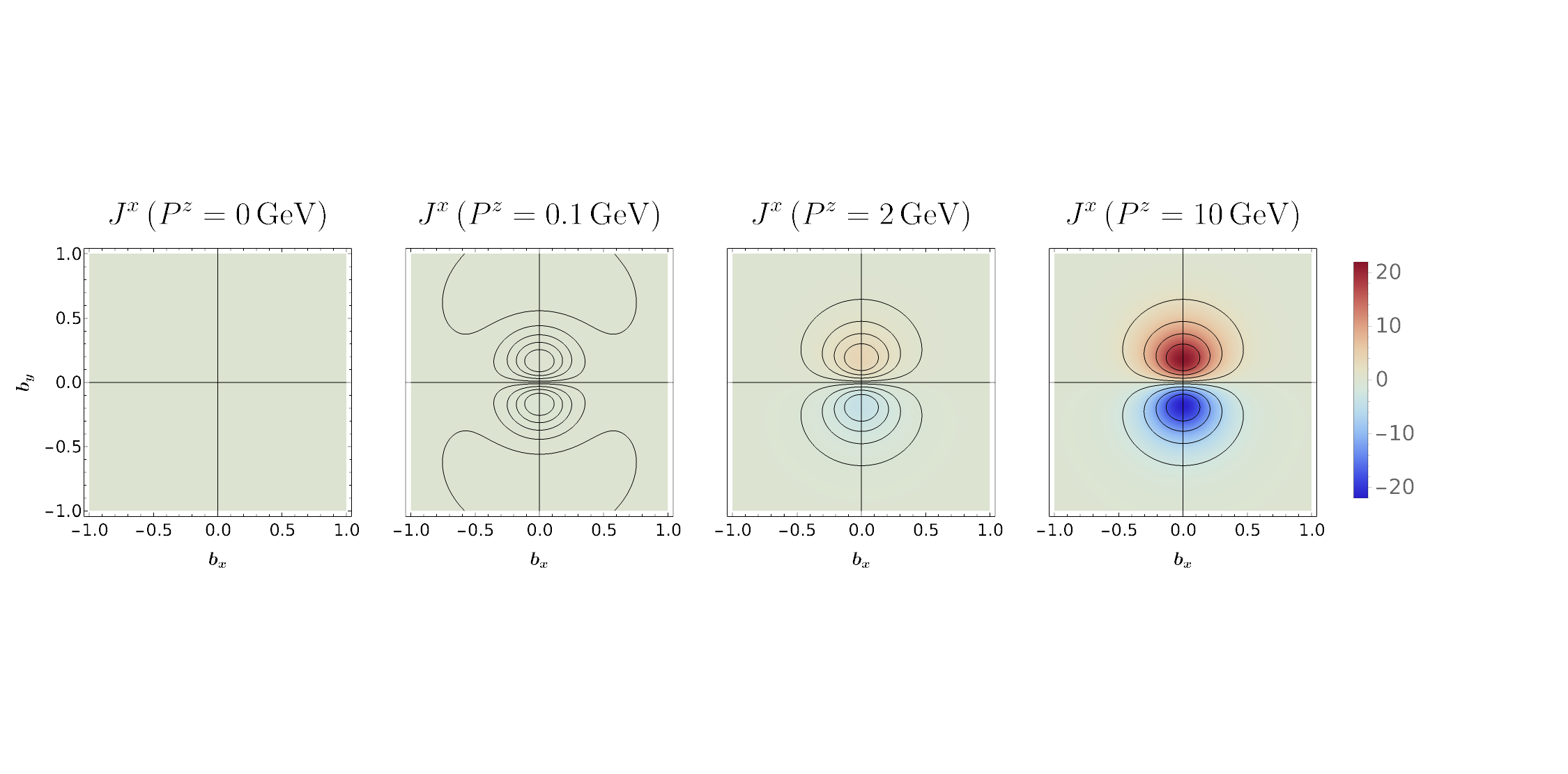}
  \hspace{0cm}

  \includegraphics[width=1\textwidth,
                 trim=0cm 4.5cm 0cm 4cm,
                 clip]{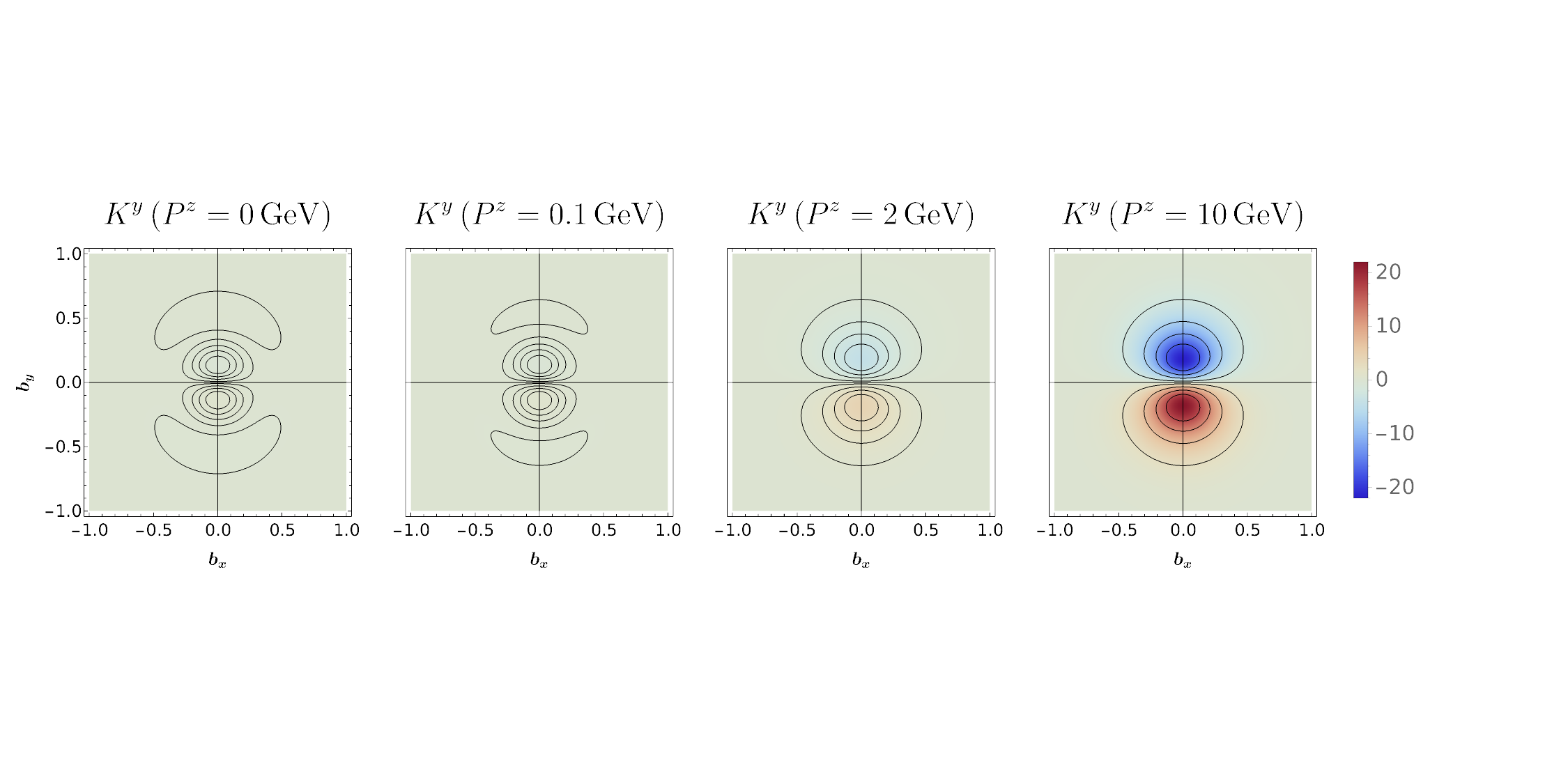}
  \hspace{0cm}

  \caption{Total (i.e.,~quark + gluon) spatial distributions of transverse TAM (upper row) and boost (lower row) in the transverse plane for a pion and different values of the pion momentum. 
    Based on the simple multipole model of Ref.~\cite{Hackett:2023nkr} for the EMT FFs, assuming a target mass of $M = 0.139$ GeV.
  }
  \label{fig:5}
\end{figure*}

We show in Fig.~\ref{fig:5} the total 2D distributions of
transverse TAM and boost for a pion in the transverse plane. From Eqs.~\eqref{Lk_spin0} and \eqref{Kk_spin0}, one can deduce that in the 2D BF, i.e., for $P^{z}=0$, the transverse TAM distribution vanishes because of the overall factor $P^{z}/M$, whereas the transverse boost distribution survives 
\begin{align}
    J^{a}
    \left(\bm{b}_{\perp},0\right)  
& = 0,\\
    K^{a}
    \left(\bm{b}_{\perp},0\right)
& = \int \frac{d^{2}\Delta_{\perp}}{\left(2\pi\right)^{2}}
    e^{-i\bm{\Delta}_{\perp}\cdot\bm{b}_{\perp}}
    i \sqrt{\tau} X_{1}^{a} \notag\\
&   \times
    4M^{2}
    \frac{d}{dt}
    \left[
  - \frac{P_{\mathrm{BF}}^{0}}{M}
    A
  + \frac{M}{P_{\mathrm{BF}}^{0}}
    F
    \right].
\label{pion_AM_distributions_BF}
\end{align}
These correspond to the first panel of TAM (upper row) and boost (lower row) in Fig.~\ref{fig:5}.

By contrast, in the IMF, i.e., for $P^{z}\to\infty$, the corresponding distributions become divergent
\begin{align}
&   \hspace{-0.5cm}J^{a}
    (\bm{b}_{\perp},\infty)  
  =  -   K^{a}
    (\bm{b}_{\perp},\infty) \notag\\
& \hspace{-0.5cm}= \int \frac{d^{2}\Delta_{\perp}}{\left(2\pi\right)^{2}}
    e^{-i\bm{\Delta}_{\perp}\cdot\bm{b}_{\perp}}
    i \epsilon_{\perp}^{ab}
    \sqrt{\tau} X_{1}^{b} \,
    4 M P^{z}
    \frac{dA}{dt}
  = \infty.
\label{pion_AM_distributions_IMF}
\end{align}
Thus, similarly to the transverse TAM distribution, the term proportional to $dA/dt$ in Eq.~\eqref{Kk_spin0} provides the leading contribution and is the only one that survives in the IMF.

Due to the dipolar structure in momentum-transfer space, the integrals of the transverse TAM and boost distributions over transverse space vanish
\begin{align}
    \hspace{-0.5cm}
    \int d^{2}b_{\perp}
    J^{a}
    (\bm{b}_{\perp},P^{z}) =
    \int d^{2}b_{\perp}
    K^{a}
    (\bm{b}_{\perp},P^{z}) 
 = 0. 
\label{pion_sum_rules}
\end{align}
This shows that, although the integrals vanish for the pion, the internal spatial structure is nontrivial. Fig.~\ref{fig:5} also suggests that the transverse boost and TAM distributions have opposite overall signs for all the considered values of $P^{z}$, and that their magnitudes become identical in the IMF limit.

\subsubsection{LF AM and boost distributions for the pion}
Similar to the TAM and boost distributions in the IF formalism, the corresponding LF distributions for a  spin-0 target are given by 
\begin{align}
    \mathcal{J}^{a}
    (\bm{b}_{\perp},P^{+})  
& = \frac{M}{P^{+}}
    \int \frac{d^{2}\Delta_{\perp}}{\left(2\pi\right)^{2}}
    e^{-i\bm{\Delta}_{\perp}\cdot\bm{b}_{\perp}}\,
    i\epsilon_{\perp}^{ab}
    \sqrt{\tau}
    X_{1}^{b}\notag\\
&   \hspace{-0.5cm}
    \times 
    \biggl[
  - 4M^{2}
    \frac{d}{dt}
    \biggl(
    \frac{P^{2}}{2M^{2}}
    A 
  - F
    \biggr)  
  + \frac{1}{2}
    D  
    \biggr],
\label{JLF_spin0_1}\\
    \mathcal{B}^{a}
    (\bm{b}_{\perp},P^{+})
& = \frac{P^+}{M} 
    \int \frac{d^{2}\Delta_{\perp}}{\left(2\pi\right)^{2}}
    e^{-i\bm{\Delta}_{\perp}\cdot\bm{b}_{\perp}}\,
    i \sqrt{\tau} 
    X_{1}^{a}\notag\\
&   \hspace{-0.5cm}
    \times
    \left(
    -4M^{2} 
    \frac{dA}{dt}
    \right).
\label{BLF_spin0_1}
\end{align}
As expected, the transverse LF TAM is a ``bad'' operator and scales as $M/P^+$, whereas the transverse LF boost is a ``good'' operator and scales as $P^+/M$. Although the spatial distributions of the transverse LF TAM and boost in the pion are non-zero, their integrals over the transverse plane vanish due to the multipole structures, analogous to the IF case.

\begin{figure*}[htbp]

  \textbf{2D spatial distributions of transverse LF TAM and boost of a pion}
   \centering
   
  \vspace{0.2cm}
		 \includegraphics[width=0.3\textwidth]{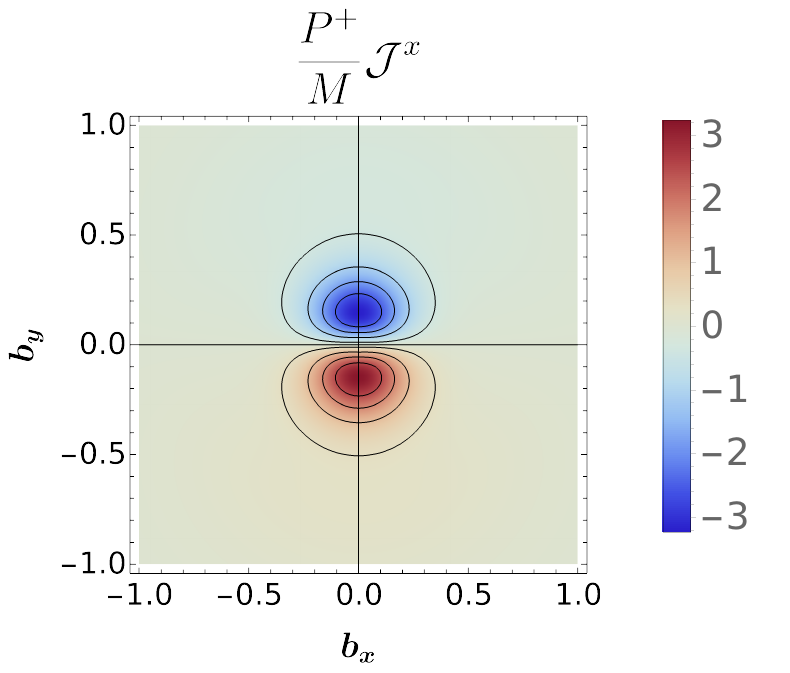}
          \hspace{0.5cm}
          \raisebox{0.13cm}{\includegraphics[width=0.3\textwidth]{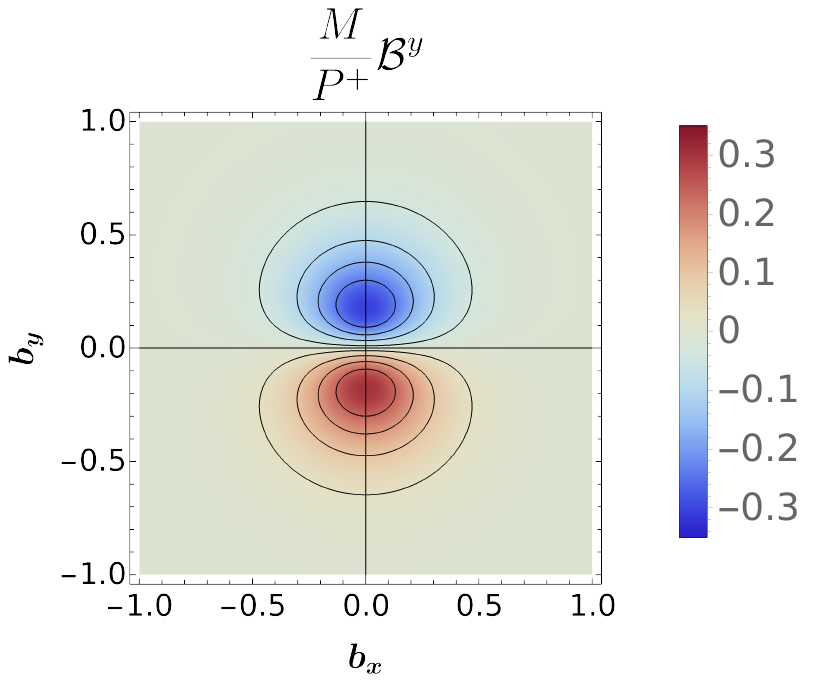}}
          \hspace{0cm}
		\caption{Total (i.e.,~quark + gluon) spatial distributions of transverse LF TAM (left panel) and boost (right panel) in the transverse plane for a pion. 
         A longitudinal LF boost factor has been included to make these distributions $P^+$-independent.
        Based on the simple multipole model of Ref.~\cite{Hackett:2023nkr} for the EMT FFs, assuming a target mass of $M = 0.139$ GeV. }
		\label{fig:6}
\end{figure*}

We illustrate in Fig.~\ref{fig:6} the total 2D distributions of transverse LF TAM and boost for a pion in the transverse plane.
These distributions are multiplied by an appropriate longitudinal LF boost factor so as to make them $P^{+}$-independent. 
In contrast to the corresponding IF distributions, these LF distributions do not exhibit a simple sign-and-magnitude relation, i.e., they are neither equal in magnitude nor related by an overall sign.
Nevertheless, it can be shown that these coincide with  the IMF limit of the pion IF TAM and boost scaled distributions given in Eqs.~\eqref{boost_EFtoLF} and \eqref{TAM_BFtoIMF}.

\section{Genuine LF spin operators}
\label{app_LF_genuine_spin_operator}

The genuine transverse spin operator in the LF framework is defined in Refs.~\cite{Harindranath:2001rc, Harindranath:2013goa} in terms of the Pauli--Lubanski operator as
\begin{equation}
    \hat{\mathscr{J}}^a = \frac{1}{M} \bigl(\hat{W}^a - \hat{P}^a \hat{\mathscr{J}}^3 \bigr), \label{def_genuine_LF_spin_operator}
\end{equation}
where 
\begin{align}
    \hat{W}^\mu = \frac{1}{2} \epsilon^{\mu \nu \alpha \beta} \hat{J}_{\nu \alpha} \hat{P}_\beta \hspace{0.5cm}  \text{and}  \hspace{0.5cm} \hat{\mathscr{J}}^3 = \frac{\hat{W}^+}{\hat{P}^+} \label{plps}
\end{align}
with $\hat{J}_{\nu\alpha}$ and $\hat{P}_\beta$ defined in Eq.~\eqref{def_AM}. Here, only the orbital contribution to $\hat{J}_{\nu\alpha}$, given in Eq.~\eqref{def_OAM_operator_densities}, is considered. Our expressions differ from those in Refs.~\cite{Harindranath:2001rc, Harindranath:2013goa} due to the conventions $n^\mu=(1,0,0,-1)/\sqrt{2}$ and $\epsilon^{+-12}=1$ used in the present work.

Expanding the transverse components of the Pauli-Lubanski pseudovector in  Eq.~\eqref{plps} and substituting in Eq.~\eqref{def_genuine_LF_spin_operator}, 
we see that the genuine LF spin operator can be written in terms of the LF OAM and boost operators as follows
\begin{align}
    \hspace{-0.5cm}
    \hat{\mathscr{J}}^{a} 
& = \frac{\hat{P}^{+}}{M} 
    \hat{\mathcal{L}}^{a} 
  - \epsilon^{ab}
    \frac{\hat{P}^{-}}{M}
    \hat{\mathcal{B}}^{b}  
  + \epsilon^{ab}
    \frac{\hat{P}^{b}}{M}
    \hat{{\mathcal{B}}}^{3}  
  - \frac{\hat{P}^{a}}{M}
    \hat{\mathscr{J}}^{3}.
\end{align}
While the transverse LF OAM $\hat{\mathcal{L}}^{a}$ and boost $\hat{\mathcal{B}}^{b}$ operators were defined in Eqs.~\eqref{def_LF_transverse_boost} and \eqref{def_LF_transverse_OAM}, the longitudinal LF boost operator is defined at $x^{+}=0$ by 
\begin{align}
    \hat{\mathcal{B}}^3 &= \hat{J}^{+-} = - \int d^{2}x_{\perp}dx^{-} \, x^{-} \hat{T}^{++}.
\end{align}
These LF operators were denoted in Refs.~\cite{Harindranath:2001rc,Harindranath:2013goa} as
\begin{align}
    \hat{\mathcal{L}}^{a}
& = \epsilon^{ab}_\perp F^{b}|_{\text{\cite{Harindranath:2001rc,Harindranath:2013goa}}}, \notag\\
    \hat{\mathcal{B}}^{a} 
& = \tilde{K}^{a}|_{\text{\cite{Harindranath:2001rc,Harindranath:2013goa}}}, 
    \qquad
    \hat{\mathcal{B}}^3   
  = \tilde{K}^{3}|_{\text{\cite{Harindranath:2001rc,Harindranath:2013goa}}}.
\end{align}
Note that $\hat{\mathscr{J}}^a$ has no explicit $x^+$-dependence in~\cite{Harindranath:2001rc, Harindranath:2013goa}, and so its expression is not affected by the restriction $x^+=0$. In the present work, throughout, we have considered  $\bm{P}_{\perp}=\bm{0}_\perp$. We then get
\begin{align}
    \hat{\mathscr{J}}^a 
   &= 
    \frac{\hat{P}^+}{M} \hat{\mathcal{L}}^a   
  - \frac{\hat{P}^-}{M}\epsilon^{ab}\hat{\mathcal{B}}^b.
\end{align}
The corresponding amplitude in the LF phase-space formalism reads
    \begin{align}
   \mathscr{J}^a(\bm{b_\perp};\lambda',\lambda)&  = \frac{P^{+}}{M}
    \mathcal{L}^{a}
    (\bm{b_\perp},P^+;\lambda',\lambda) \nonumber \\
&   
  - \epsilon^{ab} 
    \frac{P^{-}}{M}
    \mathcal{B}^{b} 
    (\bm{b_\perp},P^+;\lambda',\lambda) 
    \label{glfs_1}
\end{align}
where the momentum operator is well-localized in Wigner sense, i.e., $\langle \hat{P}^{+} \rangle_{\bm{0},\bm{P}}^{\lambda',\lambda}=P^{+}A(0)$ and $\langle \hat{P}^{-} \rangle_{\bm{0},\bm{P}}^{\lambda',\lambda}=P^{-}\left[A(0)+2\bar{C}(0)\right]$. Note that $\mathscr J^a$ does not depend on $P^+$ since we have $P^-=M^2/(2P^+)$.
\begin{figure*}[htbp]

  \textbf{2D spatial distributions of genuine LF spin of a nucleon}
          \centering
          
        \vspace{0.5cm}

          \includegraphics[width=1\textwidth,
                 trim=0cm 3.7cm 0cm 3.5cm,
                 clip]{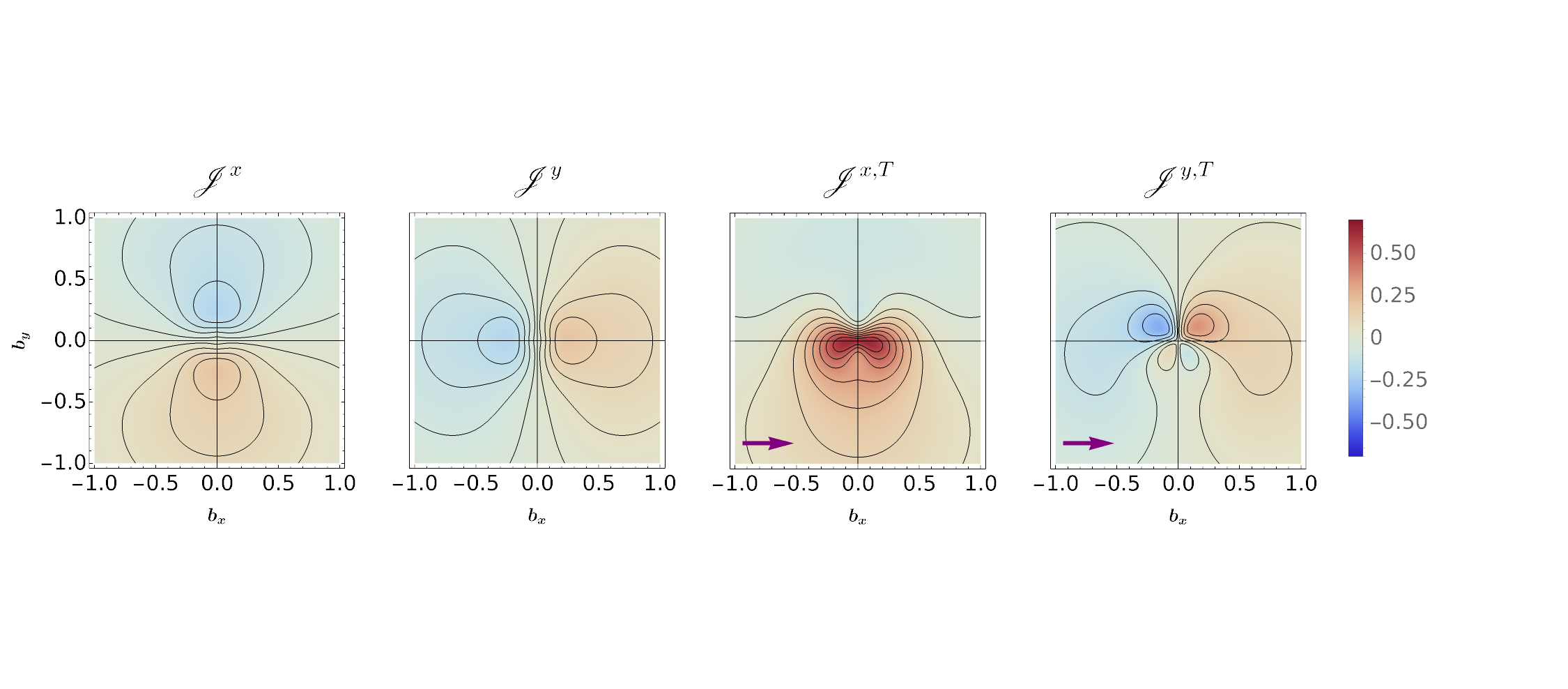}
          \hspace{0.5cm}

		\caption{Total (i.e.,~quark + gluon) spatial distributions of x and y components of genuine LF spin of Ref.~\cite{Harindranath:2001rc, Harindranath:2013goa} in the transverse plane for an unpolarized nucleon (first and second panel) and a transversely polarized nucleon along the $x$-axis (third and fourth panel). 
            Based on the simple multipole model of Ref.~\cite{Hackett:2023nkr} for the EMT FFs, assuming a target mass of $M = 0.938$ GeV.}
		\label{fig:7}
\end{figure*}

The transverse components of $\mathscr{J}^a(\bm{b_\perp};\lambda',\lambda)$ are shown in Fig.~\ref{fig:7} for both an unpolarized nucleon and a transversely polarized nucleon along the $x$-axis. The unpolarized nucleon exhibits a purely dipole structure for genuine LF spin distribution, whereas the transversely polarized nucleon receives a significant contribution from the quadrupole terms. This is different from the LF TAM distribution in our work, where quadrupole terms are much less prominent than the dipole or monopole terms.

\bibliography{PRD_TOAM}
\bibliographystyle{apsrev}

\end{document}